\documentclass[aps, prx, twocolumn, showpacs, floatfix,10pt,superscriptaddress]{revtex4-2}

\usepackage[T1]{fontenc}
\usepackage{amsmath}
\usepackage{amsfonts}
\usepackage{amsbsy}
\usepackage{amssymb}
\usepackage{graphicx}
\usepackage{textcomp}
\usepackage[caption=false,singlelinecheck=false]{subfig}
\usepackage{xcolor}
\usepackage{mathrsfs}
\usepackage{mathtools}
\usepackage{bm}
\usepackage{gensymb}
\usepackage{braket,wasysym}
\usepackage[colorlinks,linkcolor=blue,citecolor=red,filecolor=magenta,urlcolor=red,breaklinks]{hyperref}
\usepackage[bottom]{footmisc}
\usepackage{verbatim}
\hypersetup{colorlinks=true, urlcolor=blue, citecolor=red, pdfborder={0 0 0}}
\usepackage{breakurl}
\usepackage{natbib}

\allowdisplaybreaks

\begin{document}

\preprint{APS/123-QED}

\title{Multifractal collective order : Application to multifractal \\ superconductivity in dodecagonal quasicrystals}

\author{SeHyun Park}
\email{laniakea@kaist.ac.kr}
\affiliation{%
Korea Advanced Institute of Science and Technology, Daejeon 34141, South Korea\\
}%
 
\author{Junmo Jeon}%
\email{junmojeon@sophia.ac.jp}
\affiliation{Physics Division, Sophia University, Chiyoda-ku, Tokyo 102-8554, Japan}

\author{SungBin Lee}%
\email{sungbin@kaist.ac.kr}
\affiliation{%
Korea Advanced Institute of Science and Technology, Daejeon 34141, South Korea\\
}%

\date{\today}

\begin{abstract}

Conventional descriptions of collective ordering often assume spatially uniform order parameters, although the underlying electronic states can exhibit strong spatial inhomogeneity or multifractality. Here, we demonstrate that deterministic aperiodic geometry can instead stabilize multifractal collective order.
We formulate a general diagnostic applicable to superconducting, magnetic, and charge channels, combining direct scaling analysis of the order parameter with three criteria: its spatial hierarchy must be dictated by the underlying geometry, emerge in the leading ordering instability, and persist in the self-consistent ordered state below a single mean-field ordering onset. We demonstrate that superconductivity on the experimentally resolved Ti network of a dodecagonal oxide quasicrystal satisfies all three criteria. The resulting condensate exhibits multifractal scaling over the resolved range of spatial scales, retaining its hierarchical structure throughout the ordered phase and channeling flux-induced equilibrium currents along geometry-locked rings. 

\end{abstract}

\maketitle

\textit{Introduction---} A central question in inhomogeneous quantum matter is whether collective order retains the scale-dependent structure of the underlying electronic states or smooths it out through self-consistency~\cite{FeigelmanEtAl2010,FanGarciaGarcia2020,RubioVerduEtAl2020,WangEtAl2024,NikonovEtAl2025}. Since a collective order such as superconducting order parameter coherently combines contributions from many electronic states near the Fermi level~\cite{BardeenCooperSchrieffer1957,FeigelmanEtAl2010}, multifractal wave functions, known as critical states need not generate a multifractal collective ordered state; conversely, an inhomogeneous order parameter would not be multifractal~\cite{GhosalRanderiaTrivedi1998,GhosalRanderiaTrivedi2001,AraujoAndrade2019,RaiHaasJagannathan2019,RaiEtAl2020,WangEtAl2024,BiswasDebnathDutta2026}. Thus, multifractal collective order requires more than spatial modulation. We require three conditions: (i) a candidate hierarchy is fixed by the underlying geometry, independently of any electronic input; (ii) this hierarchy governs the leading extended ordering instability of the normal state; and (iii) the self-consistent order parameter retains the hierarchy and forms a single condensate with one thermodynamic onset, rather than a collection of independently condensing local states. Whether a hierarchy defined by the lattice can organize the leading extended instability and survive as multifractal order at a single onset remains an open question.

Quasicrystals can host multifractal critical states, and the Fermi level determines which states contribute most strongly to pairing~\cite{MaceEtAl2017}. Unlike Penrose and Ammann–Beenker tilings~\cite{AraujoAndrade2019}, where multifractal critical states reflect both quasiperiodic long-range order and an inhomogeneous distribution of coordination numbers, the dodecagonal Niizeki–Gähler-tiling has a uniform local coordination in the bulk. Hence, multifractality of electronic states originates primarily from the quasiperiodic geometrical hierarchy itself. This makes the system particularly suitable for realizing collective orders, including superconductivity, that inherit a common multifractal structure from critical states of the same geometric origin and thereby satisfy the three conditions above. Specifically, ultrathin BaTiO$_3$- and SrTiO$_3$-derived layers form long-range dodecagonal order on metal surfaces~\cite{ForsterEtAl2013,SchenkEtAl2017,SchenkEtAl2019,YuharaEtAl2020}, together with a family of periodic approximants~\cite{WuehrlEtAl2023}. Structural refinement of the Sr--Ti--O family identifies a two-dimensional, threefold-coordinated Ti--O network described by the Niizeki--G\"ahler tiling~\cite{SchenkEtAl2022,NiizekiMitani1987,Gaehler1988,CockayneMihalkovicHenley2016,Yamada2022}. The same titanate materials family also hosts dilute and interfacial superconductivity~\cite{SchooleyEtAl1964,ReyrenEtAl2007,CavigliaEtAl2008,CollignonEtAl2019}, with low-energy carriers derived from Ti-$3d$ orbitals~\cite{ChiangEtAl2019,CollignonEtAl2019}. This establishes a direct materials setting in which superconductivity and dodecagonal Ti networks can meet.

Before introducing any electronic structure, we identify purely geometrically a set of locally 12-fold-symmetric regions, denoted $R$, whose boundaries nearly coincide between neighboring regions. The electronic states then provide an independent test of whether this geometric hierarchy is physically active. Quasiperiodic hopping models can support critical wave functions with hierarchical spatial structure~\cite{KohmotoSutherlandTang1987,MaceEtAl2017}. The leading extended eigenmode of the normal-state Cooper susceptibility~\cite{TiceaEtAl2024,GhadimiYang2025} is concentrated on $R$, showing that the geometry organizes an extended pairing-response channel. Fully self-consistent Bogoliubov--de Gennes calculations~\cite{AraujoAndrade2019,TakemoriEtAl2020,WangEtAl2024,HoriEtAl2024,JeonLee2025,NikonovEtAl2025} then show that the superconducting order parameter inherits this hierarchy, exhibiting a broad multifractal spectrum rather than being smoothed into a homogeneous condensate. 
The multifractal structure persists to a single heat-capacity anomaly~\cite{TakemoriEtAl2020}, while a finite superfluid phase stiffness supports phase rigidity within the finite-sample mean-field calculation without restoring spatial uniformity. 
The finite phase-twist response supports mean-field phase rigidity in the finite system, despite the spatially inhomogeneous order-parameter amplitude.
Moreover, the fractal texture is not a passive amplitude pattern: under applied flux, equilibrium supercurrents circulate along the very same geometry-selected boundary rings, so the tiling dictates not only where the condensate resides but also the intrinsically ring-shaped paths of its electromagnetic response.
Thus self-consistency can transmit, rather than erase, critical single-particle structure when the hierarchy is dictated by geometry~\cite{ZhangFoster2022,FanChernLin2021,RaiEtAl2020,WangEtAl2024,NikonovEtAl2025}. Because the same diagnostic applies to magnetic~\cite{KogaTsunetsugu2017,ProfeEtAl2021} and charge~\cite{ProfeEtAl2021,LiuLiYang2025} channels, our results establish deterministic aperiodic geometry as a route to multifractality of collective quantum order itself~\cite{FanChernLin2021,WangEtAl2024,NikonovEtAl2025}, distinct from disorder-driven mechanisms~\cite{FeigelmanEtAl2007,FeigelmanEtAl2010,FanGarciaGarcia2020,RubioVerduEtAl2020}.

\begin{figure*}
\includegraphics[width=1.0\textwidth]{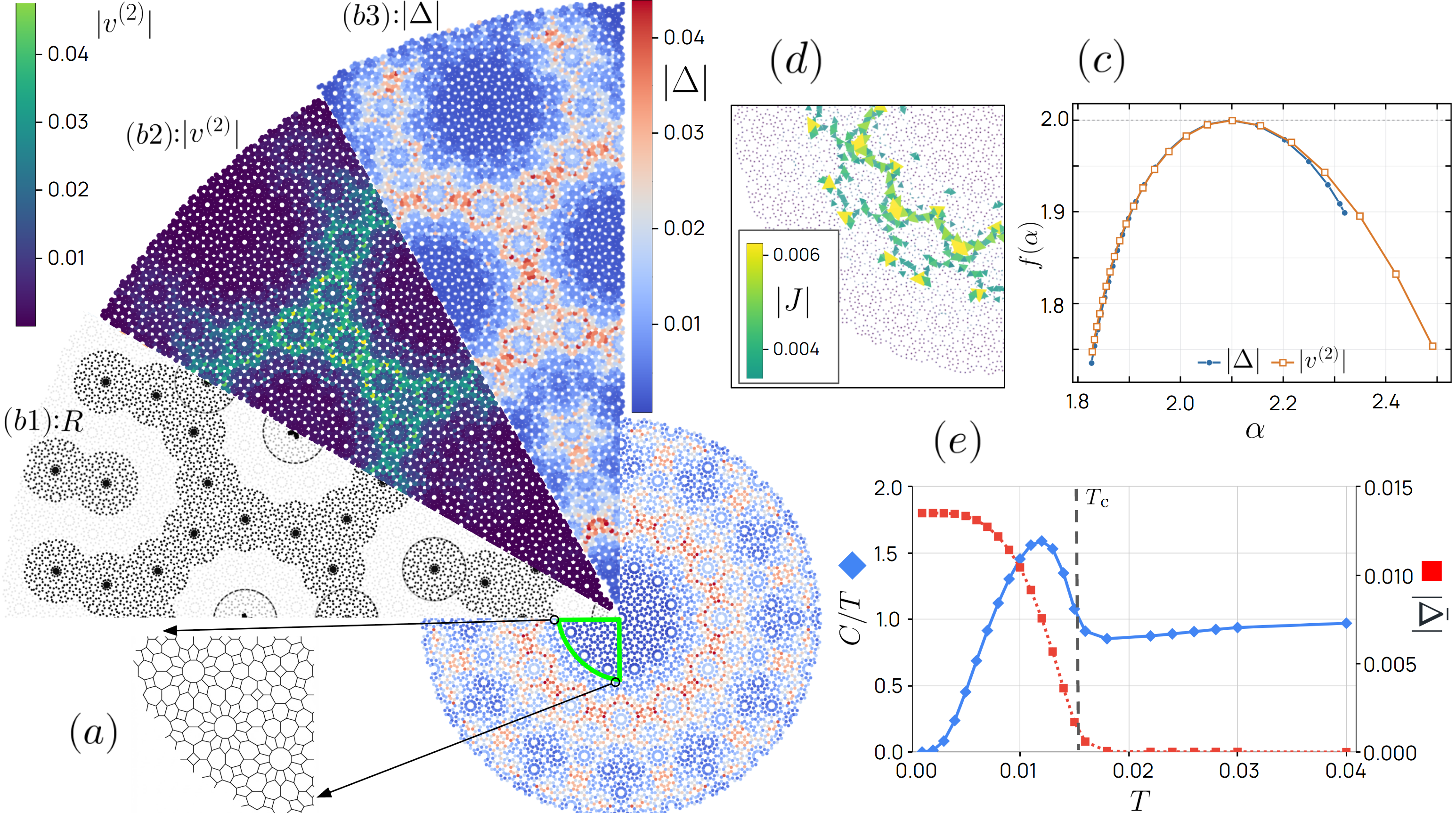}
\caption{\label{fig:main}
Multifractal superconductivity on the Ti-site network.
(a) The Ti-site network, with coordination $z=3$ at every interior site. The central disk shows the self-consistent gap $|\Delta_i|$ on a twelvefold-symmetric sample of $N=7656$ sites.
(b1)--(b3) The three-criterion diagnostic, evaluated on a $30^\circ$ sector of $5991$ sites whose two radial edges are identified by rotation.
(b1) Criterion (i): the site set $R$, the union of all locally twelvefold-symmetric patches of the Ti network, fixed by geometry alone before any electronic calculation (Appendix~\ref{app:second_radius_backbone}).
(b2) Criterion (ii): the leading extended eigenvector $|v^{(2)}_i|$ of the linearized Cooper-susceptibility kernel at $T=0.015t$
(Appendix~\ref{app:origin_multifractal_pairing}); its weight is concentrated on $R$.
(b3) Criterion (iii), first part: the self-consistent $|\Delta_i|$ at $T=0.001t$ retains the hierarchy of (b1)--(b2), as quantified in (c).
(c) Multifractal spectra $f(\alpha)$ of $|v^{(2)}_i|$ from (b2) and of $|\Delta_i|$ from (b3).
(d) Equilibrium bond supercurrents on the sample of (a) at $T=0.001t$ under an applied flux $\Phi=6.6\,\Phi_0$; arrows give the direction, color and length the magnitude.
(e) Criterion (iii), second part: the mean gap $\bar{\Delta}\equiv N^{-1}\sum_i|\Delta_i|$ and the specific heat $C_\mu/T$ versus temperature on the sample of (a). The single anomaly at $T_c\simeq0.015t$ marks a single mean-field pairing onset.
All panels: $U/t=-1$, filling $1.14$.
}
\end{figure*}

\textit{Geometric hierarchy as the candidate pattern---} Our setting is the interfacial titanate quasicrystal introduced above: an atomically thin Ba--Ti--O layer whose dodecagonal order is experimentally resolved~\cite{ForsterEtAl2013,CockayneMihalkovicHenley2016,SchenkEtAl2019,SchenkEtAl2022,Yamada2022,WuehrlEtAl2023}, belonging---together with its Sr--Ti--O siblings---to the same titanate family whose doped bulk and heterointerfaces host dilute, gate-tunable superconductivity~\cite{SchooleyEtAl1964,ReyrenEtAl2007,CavigliaEtAl2008,CollignonEtAl2019}. Ba sites form the vertices of a Niizeki--G\"ahler tiling generated by cut-and-project from the four-dimensional hypercubic lattice~\cite{NiizekiMitani1987,Gaehler1988,Yamada2022} [Appendix~\ref{app:construction}], triangle, rhombus, and square tiles are decorated with one, two, and four Ti sites, respectively, and the low-energy carriers derive from Ti $3d$ states, as established in related titanate systems~\cite{ChiangEtAl2019}. Unless stated otherwise, we perform the calculations on a $30^\circ$ sector containing $N=5991$ Ti sites, with rotationally identified angular boundaries [Appendix~\ref{app:construction}]. Hopping is defined on the Ti--Ti bonds of the network [Fig.~\ref{fig:main}(a)]; since every interior Ti site has coordination number $z=3$, no spatial structure of the order parameter can be attributed to coordination disorder.

The Ti-site network contains locally 12-fold-symmetric regions centered on selected Ba sites. The radius is denoted by $r_2$, and the geometry-defined site set $R$ (Fig.~\ref{fig:main}(b1)) is constructed from the corresponding radius-$r_2$ patches. The boundaries of these patches forms hierarchical ring patterns and, at the system size considered below, is divided into a smaller spatial region and a larger spatial region. The construction of $R$ is fixed independently of the electronic structure; further details are given in Appendix~\ref{app:second_radius_backbone}.

\textit{Pairing model---}Superconductivity in the titanate family is $s$-wave-like, carried by states of Ti $3d$ character~\cite{SchooleyEtAl1964,CollignonEtAl2019,ChiangEtAl2019}, while its microscopic pairing mechanism remains debated~\cite{CollignonEtAl2019}. We therefore make no attempt to model the pairing glue; instead we adopt the standard phenomenological weak-coupling description of a local $s$-wave channel, the onsite attractive Hubbard model on the Ti network~\cite{MicnasEtAl1990,SakaiEtAl2017,AraujoAndrade2019,TakemoriEtAl2020,LiuEtAl2022,WangEtAl2024}, retaining a single orbital per Ti site so that the results reflect the quasiperiodic geometry alone rather than orbital multiplicity:
\begin{equation}
\mathcal H=-t\sum_{\langle i,j\rangle,\sigma}
\left(\hat c_{i\sigma}^{\dagger}\hat c_{j\sigma}+\mathrm{h.c.}\right)
+U\sum_i\hat n_{i\uparrow}\hat n_{i\downarrow}
-\mu\sum_{i,\sigma}\hat n_{i\sigma},
\label{hubbard_main}
\end{equation}
where $\hat c_{i\sigma}^{\dagger}$ ($\hat c_{i\sigma}$) creates (annihilates) a spin-$\sigma$ electron at Ti site $i$, $\hat n_{i\sigma}=\hat c_{i\sigma}^{\dagger}\hat c_{i\sigma}$, and $\langle i,j\rangle$ denotes a connected Ti pair. The uniform hopping $t=1$ sets the energy unit, and the chemical potential $\mu$---the analogue of the experimentally tunable gate voltage or doping level~\cite{ReyrenEtAl2007,CavigliaEtAl2008}---selects which part of the quasicrystalline spectrum participates in pairing. The bandwidth of the non-interacting Hamiltonian is approximately $6t$, and we consider the weak onsite attractive interaction $U/t=-1$. The mean filling $\bar n$ increases by about $0.033\%$ as $T$ approaches $T_c$.

\textit{Linearized pairing diagnostic.---}Before solving the interacting model self-consistently, we characterize its bare normal-state pairing response---criterion (ii). The onsite Cooper susceptibility $\chi_{ij}(T)$ measures the pair amplitude induced at site $i$ by an infinitesimal onsite pairing field applied at site $j$~\cite{TiceaEtAl2024}. For a uniform onsite attraction $U<0$, an infinitesimal gap profile satisfies
\begin{equation}
\Delta_i=|U|\sum_j\chi_{ij}(T)\Delta_j,\qquad
\sum_j\chi_{ij}(T)v_j^{(a)}=\lambda_a(T)v_i^{(a)},
\label{linearized_gap}
\end{equation}
where $v^{(a)}$ denotes the spatial profile of the $a$th pairing-response channel, with eigenvalues ordered as $\lambda_1\geq\lambda_2\geq\cdots$, and $|U|\lambda_a=1$ gives the corresponding linearized instability condition~\cite{TiceaEtAl2024,GhadimiYang2025}. Because distinct response channels may be concentrated in different spatial regions, we examine their eigenmode profiles relative to the preselected domain $R$, while keeping $R$ fixed independently of $\chi_{ij}$. The explicit expression for $\chi_{ij}$ and the construction of its eigenmodes are given in Appendix~\ref{app:cooper_susceptibility}. This bare normal-state response provides a spatial reference for the self-consistent calculation below.

\textit{Self-consistent order and thermodynamics.---}In the weak-coupling regime, where the interaction strength is sufficiently small compared with the bandwidth, we solve the Hamiltonian in Eq.~\eqref{hubbard_main} within the mean-field approximation~\cite{BardeenCooperSchrieffer1957,MicnasEtAl1990,Zhu2016}. Without external field, the mean-field Hamiltonian is decoupled in the onsite spin-singlet pairing and spin-balanced Hartree channels~\cite{GhosalRanderiaTrivedi2001,JeonLee2025,NikonovEtAl2025} and solved self-consistently with the Bogoliubov--de Gennes (BdG) equations~\cite{Zhu2016,AraujoAndrade2019,HoriEtAl2024,JeonLee2025,NikonovEtAl2025}. We define the local order parameter as $\Delta_i=U\langle\hat c_{i\downarrow}\hat c_{i\uparrow}\rangle$ and the Hartree potential as $V_i^{\rm H}=Un_i/2$, where $n_i=\sum_\sigma\langle\hat n_{i\sigma}\rangle$ is the local density; the BdG equations, self-consistency conditions, and thermodynamic expressions are given in Appendix~\ref{app:bdg}. The Hartree--BdG equations are solved self-consistently at each sampled temperature, including both $\Delta_i$ and the spatially varying $V_i^{\rm H}$, with the order parameter classified as numerically zero when $|\Delta_i|<10^{-6}$. The entropy $S(T)$ is evaluated from the converged BdG spectrum, and the heat capacity $C(T)=T\,dS(T)/dT$~\cite{MicnasEtAl1990,Zhu2016,JeonLee2025} locates the thermodynamic pairing onset required by criterion (iii).

\textit{Multifractal condensate---}In the local tight-binding calculation described in Appendix~\ref{app:second_radius_backbone}, an eigenstate near filling $1.14$ has substantial weight near the radius-$r_2$ boundary. With the geometry, the linearized diagnostic, and the self-consistent machinery in place, we now turn to the central result. Figure~\ref{fig:main}(b3) shows a self-similar spatial structure of the order parameter, comprising hierarchical ring patterns across multiple length scales. At the system size considered here, $R$ is divided into two spatial regions. The eigenvector associated with the largest eigenvalue, $v^{(1)}$, is predominantly localized in the smaller region, whereas $v^{(2)}$, associated with the second-largest eigenvalue, has the highest eigenvalue among the modes extending across the larger region. Quantitatively, $97.57\%$ of the normalized weight of $v^{(1)}$ lies in the smaller inner region and $0.018\%$ in the larger extended region; the corresponding weights for $v^{(2)}$ are $0.323\%$ and $93.41\%$, respectively. The self-consistent $|\Delta_i|$ has weight in both regions. Since only the larger region spans multiple length scales, we use $v^{(2)}$ to characterize its multifractal distribution. The extended structures of $v^{(2)}$ and $|\Delta_i|$ both follow the radius-$r_2$ patches defining $R$, indicating their common geometric organization, compare Figs.~\ref{fig:main}(b1)--(b3). 

In an interfacial realization, this texture motivates a spectroscopic test: spatially resolved gap spectroscopy co-registered with the lattice~\cite{PanEtAl2001,GomesEtAl2007,RubioVerduEtAl2020} images the local spectral gap, which can reflect the underlying pairing texture, although it need not coincide locally with $|\Delta_i|$; because the ring hierarchy extends over multiple length scales, the geometry-locked multiring pattern provides a concrete target for spatially resolved spectroscopy.
 
To quantify this shared organization, we apply standard box-counting multifractal analysis~\cite{HalseyEtAl1986,EversMirlin2008} to the normalized spatial measures of $v^{(2)}$ and $\Delta_i$: the mass exponent $\tau_X(q)$ ($X=v^{(2)},\Delta$) characterizes the scale dependence of the $q$th moment, with positive (negative) $q\in (-1.2,3.2)$ emphasizing regions of large (small) spatial weight, and the multifractal spectrum follows from the Legendre transformation $\alpha_X(q)=d\tau_X(q)/dq$, $f_X(\alpha)=q\alpha_X(q)-\tau_X(q)$ [Appendix~\ref{app:multifractal_spectrum}]. The resulting spectra, Fig.~\ref{fig:main}(c), are broad for both $\Delta_i$ and $v^{(2)}$---consistent with multifractal scaling over the resolved range of spatial scales---and nearly coincide at small $\alpha$, while they deviate at large $\alpha$. Because the two measures use different spatial supports, the present comparison is qualitative. Thus, the self-consistent order parameter retains the geometry-selected extended hierarchy while also carrying weight in the compact inner region.

\textit{Geometry-selected currents and phase stiffness.---} The flux response and the phase stiffness are computed on a smaller circular sample that keeps the 12-fold symmetry but is treated without the sector reduction [Appendix~\ref{app:supercurrent}]. To calculate the equilibrium current response to an applied flux~\cite{SakaiArita2019,Nagai2022,FukushimaEtAl2023}, we thread a pure-gauge flux line through the sample origin, $\mathbf{A}(x,y)=3.3\left(-\frac{y}{x^2+y^2},\frac{x}{x^2+y^2}\right)$, incorporated into the hopping amplitudes through the Peierls substitution~\cite{Peierls1933,GrafVogl1995}; for a bond connecting sites $j$ and $i$, the Peierls phase is $\phi_{ij}=-\frac{e}{\hbar}\int_{\mathbf r_j}^{\mathbf r_i}\mathbf A\cdot d\mathbf l$. With this choice the magnetic field vanishes on every bond, while loops encircling the origin acquire a gauge-invariant Aharonov--Bohm phase. The enclosed flux, $\Phi=3.3\,h/e=6.6\,\Phi_0$ with $\Phi_0=h/2e$, is a non-integer multiple of the superconducting flux quantum, so a phase-coherent condensate cannot remove the phase twist by a single-valued gauge choice and must respond with a persistent equilibrium current [Appendix~\ref{app:supercurrent}]. The external vector potential is held fixed during the self-consistent calculation, and the equilibrium current on each bond is evaluated from the corresponding one-particle density matrix. The reported current is the total equilibrium BdG bond current in the presence of the applied vector potential. At the same applied flux and model parameters, the equilibrium current is zero when $\Delta_i=0$, and hence no normal-state current is subtracted. This finite-flux current map is used to determine the spatial paths of the current and is not used as a direct measure of the phase stiffness~\cite{ScalapinoWhiteZhang1992,ScalapinoWhiteZhang1993,SchirmerRossi2025}. Separately, we evaluate the temperature-dependent bare phase stiffness $\rho_s(T)$, neglecting renormalization by vortex--antivortex fluctuations. Details of the bond-current and phase-stiffness calculations are provided in Appendices~\ref{app:supercurrent} and~\ref{app:bkt_estimate}, respectively.

The hierarchical pairing texture persists up to a single mean-field pairing onset. The heat capacity exhibits a single anomaly near $T_c\simeq0.015t$, where the spatially averaged amplitude $\bar{\Delta}$ vanishes within numerical resolution [Fig.~\ref{fig:main}(e)]. As the temperature increases, the pairing amplitudes in both the inner ring and the extended hierarchical region decrease toward this common onset [Appendix~\ref{app:bdg}], consistent with a single self-consistent pairing texture rather than separate local pairing transitions. Self-consistent weak-probe calculations yield a finite mean-field phase-stiffness scale of order $10^{-3}t$ at low temperatures, decreasing over the sampled temperature range [Appendix~\ref{app:bkt_estimate}]. This finite response supports phase rigidity in the finite system despite the strongly inhomogeneous pairing amplitude.

Under an applied flux, the hierarchical multifractal superconducting state confines the induced current predominantly to the ring structures where the order parameter is concentrated, forming a network of ring-shaped circulating current paths within the bulk [see Fig.~\ref{fig:main}(d)]. This spatial correspondence indicates that the multifractal structure is not merely an amplitude modulation of the superconducting order parameter, but directly organizes the pathways of the electromagnetic response. The geometrically generated hierarchy of the condensate can therefore be imprinted on its current-carrying structure, providing a direct connection between the multifractal superconducting texture and its observable response to an external gauge field.

\textit{Conclusion---}Our results show that self-consistency need not erase the spatial hierarchy imposed by deterministic aperiodic geometry. On the dodecagonal Ti-site network, the superconducting order parameter exhibits multifractal scaling over the resolved range of spatial scales. A structural pattern defined solely by the tiling appears in the leading extended mode of the bare normal-state Cooper response and is retained by the self-consistent condensate. The inner and extended components approach a common mean-field pairing onset, while weak-probe calculations support mean-field phase rigidity in the finite system despite the strongly inhomogeneous amplitude. This correspondence motivates a general diagnostic for geometry-organized multifractal collective order: identify a candidate spatial hierarchy independently of the ordered state, examine its expression in the normal-state response, and test whether it survives self-consistency in an ordered state whose spatial components share a single mean-field ordering onset. Because these steps are not specific to superconductivity, the same framework can be extended to other collective channels.

The experimentally established structures of dodecagonal titanate layers provide a concrete setting for testing these predictions~\cite{ForsterEtAl2013,SchenkEtAl2019,SchenkEtAl2022}. If pairing can be induced by gating, doping, or proximity coupling~\cite{ReyrenEtAl2007,CavigliaEtAl2008,WangEtAl2012FeSe}, spatially resolved gap spectroscopy co-registered with the lattice~\cite{PanEtAl2001,GomesEtAl2007,RubioVerduEtAl2020,RanderiaEtAl2016,HamidianEtAl2016}, interpreted alongside calculated local spectra, could test for modulations correlated with the $R$-boundary network and their temperature evolution. Comparisons with periodic approximants~\cite{YuharaEtAl2020,WuehrlEtAl2023}, complemented by numerical studies of phason rearrangements and disorder, would probe the robustness of the geometry-selected hierarchy. Geometry-locked currents offer a complementary target in artificial quasicrystalline superconducting networks~\cite{BehroozEtAl1986,SpringerVanHarlingen1987,ItzlerEtAl1993}, while atomic quasicrystalline superconductors~\cite{KamiyaEtAl2018,ShiinoEtAl2021,TokumotoEtAl2024,MeenaEtAl2025,MatsudairaEtAl2026} provide further platforms for these diagnostics. An important theoretical extension is to connect this real-space organization to the quantum geometry of the electronic states. Recent work has found that the zero-temperature superfluid weight of quasiperiodic systems hosting critical states is dominated by its quantum-geometric part~\cite{SaitoEtAl2026}.

Beyond this specific realization, our results suggest that deterministic aperiodicity can be used to control the spatial organization of collective order. Its motifs provide a means to influence where ordering tendencies are enhanced, how high-weight regions connect across scales, and which paths carry the response to external fields~\cite{SakaiEtAl2017,GhadimiYang2025,JeonLee2025,BiswasDebnathDutta2026,RaiEtAl2020,WangEtAl2024,NikonovEtAl2025,FukushimaEtAl2023,SandbergEtAl2024}. The same strategy could be pursued for magnetic and charge order~\cite{KogaTsunetsugu2017,ProfeEtAl2021,LiuLiYang2025}, and for other collective channels whose ordering tendencies can be tuned while the geometric network remains fixed~\cite{CaoEtAl2020,KhosravianLado2021,SakaiArita2019,GhadimiEtAl2021,HoriEtAl2024,KobialkaEtAl2024}. This perspective naturally suggests an inverse-design problem~\cite{Zunger2018,BedollaMontielEtAl2024}: specify a target collective texture, search over admissible aperiodic geometries, rank candidates using susceptibility eigenvalues and spatial profiles, and validate the selected textures and responses self-consistently. In this view, quasicrystalline geometry is not merely a host for unusual electronic states, but a design resource for collective quantum matter.

\begin{acknowledgments}
We thank YouYoung Joung and Jemin Park for a careful reading of the manuscript and detailed, constructive feedback. We also thank Kazuhiko Deguchi for fruitful discussions on the experimental relevance of this work.

This work was supported by  National Research Foundation of Korea (\mbox{RS-2026-00281839}, \mbox{RS-2023-00281839}) and Air Force Office of Scientific Research under FA23862514054. J.J was supported by JSPS KAKENHI Grant No. JP25H01397, JP25H01398, JP25K24854, and JP25K24855.
\end{acknowledgments}

\bibliographystyle{apsrev4-2}
\bibliography{Reference}

\newpage
\appendix
\renewcommand{\thefigure}{A\arabic{figure}}
\setcounter{figure}{0}

\section{Quasicrystal Construction and Ti Dual Lattice}
\label{app:construction}

The dodecagonal Niizeki--G\"ahler tiling is generated by a cut-and-project construction from $\mathbb Z^4$.  For $\mathbf n=(n_1,n_2,n_3,n_4)^T\in\mathbb Z^4$, we define
\begin{equation}
\mathbf r_\parallel = Q_\parallel \mathbf n, \qquad \mathbf r_\perp = Q_\perp \mathbf n , \label{cut_project}
\end{equation}
with
\begin{align} 
Q_\parallel &= \frac{a_d}{\sqrt{6}}
\begin{pmatrix} \sqrt{3} & 2 & 0 & -1 \\ -1 & 0 & 2 & \sqrt{3} \end{pmatrix}, \label{Q_parallel_app} \\
Q_\perp &= \frac{a_d}{\sqrt{6}}
\begin{pmatrix} -\sqrt{3} & 2 & 0 & -1 \\ -1 & 0 & 2 & -\sqrt{3} \end{pmatrix}. \label{Q_perp_app}
\end{align}
Here $a_d$ fixes the overall length scale.  The convention used in the main text corresponds to setting $a_d=1$.

The perpendicular-space acceptance domain is the regular dodecagon
\begin{equation}
\mathcal W = \operatorname{int}\operatorname{conv} \left\{ R_{\mathcal W}
\begin{pmatrix} \cos(k\pi/6)\\ \sin(k\pi/6) \end{pmatrix}
\ \middle|\ k=0,\ldots,11 \right\},
\label{acceptance_window_app}
\end{equation}
where $R_{\mathcal W}=0.95$. It is centered at the perpendicular-space origin, with one vertex oriented along the positive perpendicular-space $x$ axis. The Ba-site vertex set is then
\begin{equation}
\mathcal V_{\rm Ba} = \left\{ Q_\parallel\mathbf n \ \middle|\ \mathbf n\in\mathbb Z^4,\; Q_\perp\mathbf n\in\mathcal W \right\}.
\label{Ba_vertex_set_app}
\end{equation}
For the finite calculations, accepted vertices are retained within $|\mathbf r_\parallel|<R_c$, with $R_c=70$. The electronic calculations are performed on a $30^\circ$ sector. During graph construction, a narrow angular buffer is used to identify hopping bonds crossing the angular edges under a $30^\circ$ rotation. These edges are identified without an additional phase twist, while the outer radial boundary remains open. The results obtained with this construction refer to the sector graph with the specified angular boundary condition.

The Ti-site network is obtained by decorating each triangular, rhombic, and square tile with one, two, and four Ti sites, respectively. The Ti positions are fixed by affine combinations of the cyclically ordered Ba vertices $\mathbf v_k$:
\begin{align}
\boldsymbol\tau_{\triangle} &=
\frac{\mathbf v_1+\mathbf v_2+\mathbf v_3}{3}, \label{triangle_ti_position_app}
\\
\boldsymbol\tau_{\lozenge,1} &= 0.35\mathbf v_1+0.65\mathbf v_3, \  \boldsymbol\tau_{\lozenge,2} = 0.65\mathbf v_1+0.35\mathbf v_3, \label{rhombus_ti_positions_app}
\\
\boldsymbol\tau_{\square,k} &= \frac{1-\eta}{2} (\mathbf v_k+\mathbf v_{k+1}) + \frac{\eta}{2} (\mathbf v_{k+2}+\mathbf v_{k+3}), \label{square_ti_positions_app}
\\
\eta &= \frac{1}{4+2\sqrt3},
\end{align}
where $\mathbf v_1\mathbf v_3$ denotes the long diagonal of the rhombus and the square indices are understood modulo four. The coefficients $0.35$ and $\eta=1/(4+2\sqrt{3})$ specify the idealized Ti decoration used in the present lattice construction.

In the Ba--Ti--O structure, hopping between Ti-derived states is mediated by the intervening O sites. The O orbitals are not retained explicitly in the present single-orbital model; their role is incorporated into an effective Ti-only hopping graph. Let $\ell_{\rm Ba}$ denote the fundamental Ba-edge length and define the characteristic Ti-bond scale
\begin{equation}
d_{\rm Ti} = \frac{\ell_{\rm Ba}}{2.1}, \qquad \delta d_{\rm Ti} = 0.27d_{\rm Ti}.
\label{ti_bond_scale_app}
\end{equation}
Two Ti sites are connected when their separation belongs to this bond shell,
\begin{equation}
A_{ij} =
\begin{cases}
1, & \left| |\boldsymbol\tau_i-\boldsymbol\tau_j| -d_{\rm Ti} \right| < \delta d_{\rm Ti}, \\
0, & \text{otherwise}.
\end{cases}
\label{adjacency_app}
\end{equation}
The resulting adjacency matrix defines the uniform-hopping Hamiltonian used below.

The angular boundaries of the computational sector are identified by the 12-fold rotational construction, whereas the outer radial boundary is not passivated or coordination corrected. The coordination of each retained Ti site is therefore determined directly from
\begin{equation}
z_i=\sum_jA_{ij}.
\label{coordination_app}
\end{equation}
All interior Ti sites have $z_i=3$; sites with $z_i=2$ or $z_i=1$ occur only along the open outer boundary. Boundary physics is not analyzed in this work.

\section{Self-Consistent BdG Equations and Thermodynamics}
\label{app:bdg}

We use the attractive Hubbard model on the Ti dual lattice,
\begin{equation}
H = -\sum_{\langle i,j\rangle,\sigma}
(t_{ij}e^{i\phi_{ij}} c_{i\sigma}^{\dagger}c_{j\sigma}+\mathrm{h.c.})
-\mu\sum_{i,\sigma}n_{i\sigma} +U\sum_i n_{i\uparrow}n_{i\downarrow}.
\label{app_hubbard}
\end{equation}
Here $c_{i\sigma}^{\dagger}$ creates an electron with spin $\sigma$ at site $i$, $n_{i\sigma}=c_{i\sigma}^{\dagger}c_{i\sigma}$, and $\phi_{ij}$ is the Peierls phase. In zero field, $\phi_{ij}=0$. The calculations in the main text use $U/t=-1$, $\mu=-0.23t$, and $t=1$.

The Hamiltonian is defined on the finite Ti graph constructed in Appendix~\ref{app:construction}. Its angular boundaries are already identified by the 12-fold rotational construction, while the outer radial boundary remains open; no additional boundary bonds or boundary potentials are introduced in the BdG calculation.

For onsite spin-singlet pairing, we define
\begin{equation}
F_i=\langle c_{i\downarrow}c_{i\uparrow}\rangle,
\qquad
\Delta_i = UF_i = -U\langle c_{i\uparrow}c_{i\downarrow}\rangle .
\label{app_gap_definition}
\end{equation}

The Hartree shifts are determined from the opposite-spin densities,
\begin{equation}
V_{i\uparrow}^{\rm H} = Un_{i\downarrow}, \qquad V_{i\downarrow}^{\rm H} = Un_{i\uparrow}.
\label{app_hartree_spin_resolved}
\end{equation}
The converged solutions considered here are spin balanced, $n_{i\uparrow}=n_{i\downarrow}=n_i/2$, so both shifts reduce to
\begin{equation}
V_i^{\rm H} = \frac{U}{2}n_i, \qquad n_i = \sum_\sigma
\langle n_{i\sigma}\rangle .
\label{app_hartree_field}
\end{equation}
The corresponding single-particle Hamiltonian is
\begin{equation}
h_{ij}^{\rm H} = -tA_{ij}e^{i\phi_{ij}} + \left( V_i^{\rm H}-\mu_{\rm BdG} \right)\delta_{ij}.
\label{app_hartree_single_particle}
\end{equation}

The finite-temperature BdG calculations are grand canonical: $\mu_{\rm BdG}$ is held fixed, while the filling
\begin{equation}
\bar n(T) = \frac{1}{N} \sum_i n_i(T)
\label{app_filling}
\end{equation}
is obtained from the self-consistent solution and may vary weakly with temperature. Here $N$ denotes the number of Ti sites. At the reference temperature $T_{\rm ref}=0.001t$, $\mu_{\rm BdG}=-0.23t$ gives $\bar n=1.14$.
For the bare hopping problem, $\mu_0=0.4t$ is chosen by matching its particle number to the same reference filling,
\begin{equation}
n_0(\mu_0,T_{\rm ref}) = n_{\rm BdG}(\mu_{\rm BdG},T_{\rm ref}).
\label{app_mu_matching}
\end{equation}
This matching equates the mean filling at $T_{\rm ref}$, but does not equate the bare and Hartree-shifted single-particle spectra or their spatial density profiles.

The mean-field Hamiltonian can then be written as
\begin{align}
H_{\rm MF} &= \sum_{i,j,\sigma} c_{i\sigma}^{\dagger}h_{ij}^{\rm H}c_{j\sigma}
+\sum_i \left( \Delta_i c_{i\uparrow}^{\dagger}c_{i\downarrow}^{\dagger} +\Delta_i^*c_{i\downarrow}c_{i\uparrow} \right) \nonumber\\
&\quad -\sum_i\frac{|\Delta_i|^2}{U} -\frac{U}{4}\sum_i n_i^2 .
\label{app_mean_field_hamiltonian}
\end{align}
The last two terms are the mean-field constants associated with the pairing and Hartree decouplings.

Using the Nambu spinor
\begin{equation}
\Psi_n(i)= \begin{pmatrix} u_n(i)\\ v_n(i) \end{pmatrix}, \label{app_nambu_spinor}
\end{equation}
the BdG eigenvalue equation is
\begin{equation}
\sum_j
\begin{pmatrix}
h_{ij}^{\rm H} & +\Delta_i\delta_{ij}\\
+\Delta_i^*\delta_{ij} & -h_{ij}^{{\rm H}*}
\end{pmatrix}
\begin{pmatrix} u_n(j)\\ v_n(j) \end{pmatrix}
= E_n \begin{pmatrix} u_n(i)\\ v_n(i) \end{pmatrix}.
\label{app_bdg_matrix}
\end{equation}
The BdG spectrum has the usual Nambu particle-hole redundancy. In the self-consistency equations below, the sum is taken over positive eigenvalues $E_n>0$.

The local gap and density are updated as
\begin{align}
\Delta_i &= -U\sum_{E_n>0} u_n(i)v_n^*(i) \left[1-2f(E_n)\right], \label{app_gap_update} \\
n_i &= 2\sum_{E_n>0} \left[ |u_n(i)|^2 f(E_n) + |v_n(i)|^2\{1-f(E_n)\} \right], \label{app_density_update}
\end{align}
where
\begin{equation}
f(E)=\frac{1}{e^{E/T}+1}. \label{app_fermi_function}
\end{equation}
We set $k_B=\hbar=1$. 

The pairing field is updated using linear mixing with $\alpha=0.01$. Because the density converges faster than the pairing field, convergence is determined from $\max_i|\Delta_{i,\mathrm{raw}}^{(k+1)}-\Delta_i^{(k)}|<\varepsilon_\Delta$. At termination, the density residual $\max_i|n_i^{(k+1)}-n_i^{(k)}|$ remains below $\varepsilon_n$ for all sampled temperatures. We set $\varepsilon_\Delta=\varepsilon_n=10^{-5}.$

For a converged solution, the grand potential is
\begin{align}
\Omega &= -2T\sum_{E_n>0}\ln\left[2\cosh\left(\frac{E_n}{2T}\right)\right] 
+\sum_i\left(V_i^{\rm H}-\mu_{\rm BdG}\right) \nonumber\\
&\quad -\sum_i\frac{|\Delta_i|^2}{U} -\frac{U}{4}\sum_i n_i^2 . \label{app_grand_potential}
\end{align}

The entropy is computed from the positive-energy quasiparticle spectrum,
\begin{equation}
\begin{aligned}
S(T) ={}& -2\sum_{E_n>0} \left[ f(E_n)\ln f(E_n) \right. \\
&\qquad\left. +\{1-f(E_n)\}\ln\{1-f(E_n)\} \right].
\end{aligned} \label{app_entropy}
\end{equation}
where the prefactor accounts for the spin degeneracy when a single spin block is diagonalized. The heat capacity is obtained from
\begin{equation}
C_\mu(T)=T\left(\frac{\partial S}{\partial T}\right)_{\mu_{\rm BdG}}. \label{app_heat_capacity}
\end{equation}
The entropy is evaluated from a separately converged BdG solution at each sampled temperature, with $\mu_{\rm BdG}$ held fixed. The derivative is evaluated by a centered finite difference with a step of $0.001t$.

\begin{figure}
\includegraphics[width=0.95\columnwidth]{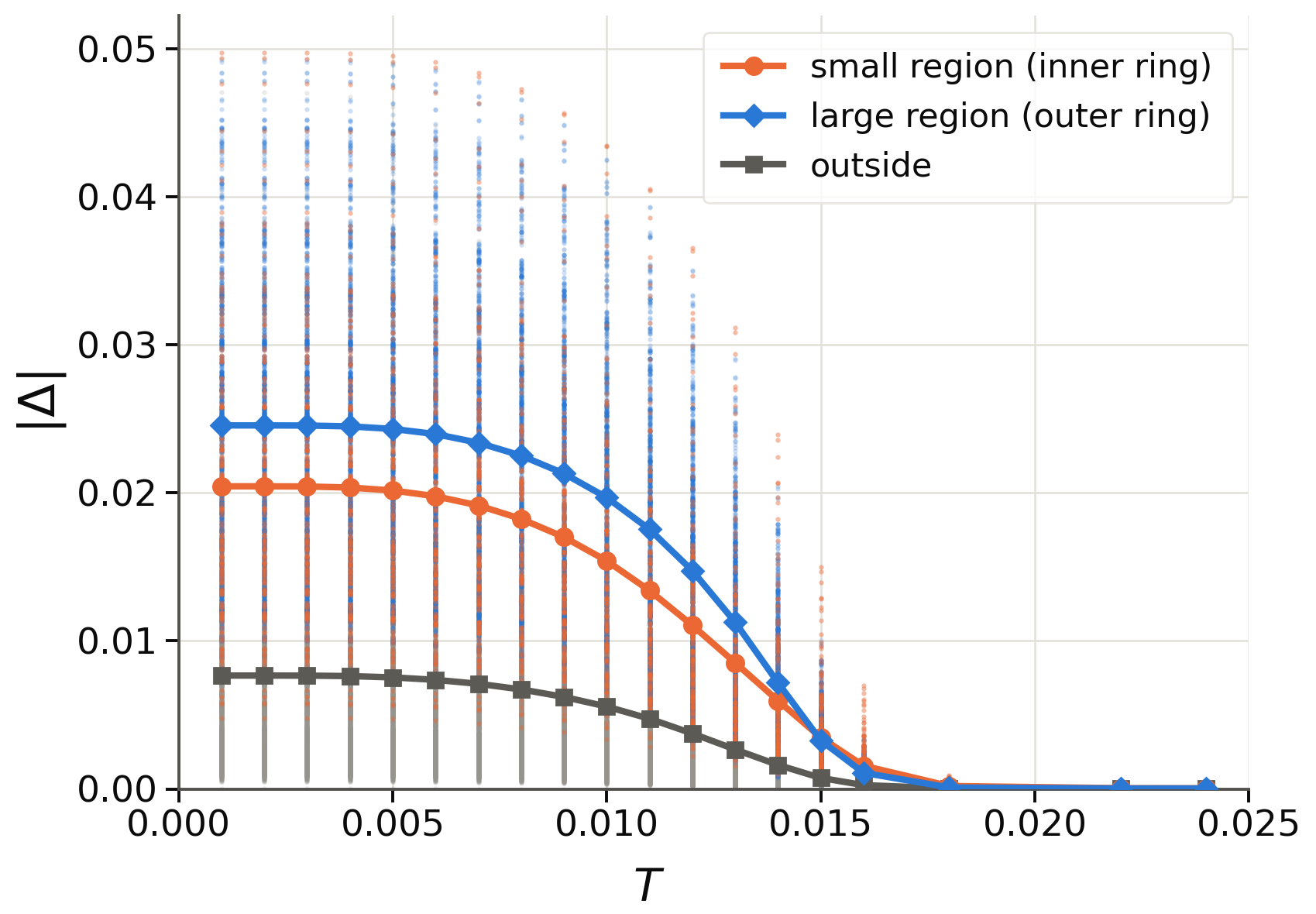}
\caption{
\label{fig:heat_app}
Temperature evolution of the site-resolved superconducting amplitude $|\Delta_i|$ for $|U|/t=1$, fixed $\mu_{\rm BdG}/t=-0.23$, and $N=5991$. Blue, orange, and gray markers denote sites in the extended component of $R$ (Appendix~\ref{app:second_radius_backbone}), the inner localized ring, and the complement of $R$, respectively. The sampled temperature dependence is compared with the mean-field onset identified by the heat-capacity anomaly.}
\end{figure}

Figure~\ref{fig:heat_app} shows the temperature evolution of the distribution of $|\Delta_i|$. The amplitudes decrease with increasing temperature, consistent with a single numerically resolved mean-field pairing onset near $T_c\simeq0.015t$. Below $T_c$ the overall amplitude decreases [Fig.~\ref{fig:main}(e)], while the multifractal form of the distribution is retained.

\section{Cooper Susceptibility and the Linearized Gap Equation}
\label{app:cooper_susceptibility}


The normal-state origin of the superconducting texture is diagnosed using the onsite Cooper-pair susceptibility. The onsite anomalous amplitude is
\begin{equation}
F_i=\langle c_{i\downarrow}c_{i\uparrow}\rangle ,
\label{app_anomalous_amplitude}
\end{equation}
and the local gap is $\Delta_i=UF_i$ with $U<0$. 

To distinguish the self-consistent gap from the external field used to define the response, we introduce a weak static pair source $\eta_i$ through
\begin{equation}
H_\eta = -\sum_i \left(\eta_i c_{i\uparrow}^{\dagger}c_{i\downarrow}^{\dagger} + \eta_i^*c_{i\downarrow}c_{i\uparrow}\right).
\end{equation}

Close to the superconducting instability, the response of the normal state to a weak static pair source is linear:
\begin{align}
F_i = \sum_j \chi_{ij}(T)\eta_j+O(\eta^3).
\label{app_linear_response} \\
\chi_{ij}(T) = \int_0^\beta d\tau\,
\left\langle T_\tau
c_{i\downarrow}(\tau)c_{i\uparrow}(\tau)
c_{j\uparrow}^{\dagger}(0)c_{j\downarrow}^{\dagger}(0)
\right\rangle_0 .
\label{app_pair_susceptibility_def}
\end{align}
The mean-field pairing term in Eq.~\eqref{app_mean_field_hamiltonian} corresponds to $\eta_i=-\Delta_i$.
The expectation value in Eq.~\eqref{app_pair_susceptibility_def} is evaluated in the normal state. The linearized gap equation is therefore
\begin{equation}
\Delta_i
=
-U\sum_j\chi_{ij}(T)\Delta_j.
\label{app_linearized_gap}
\end{equation}

Let $\phi_a(i)$ be the eigenfunctions of the normal-state single-particle Hamiltonian,
\begin{equation}
\sum_j h_{ij}^{(0)}\phi_a(j)=\epsilon_a\phi_a(i),
\qquad
\xi_a=\epsilon_a-\mu_0 .
\label{app_normal_eigenproblem}
\end{equation}
Here $h_{ij}^{(0)}=-tA_{ij}$ denotes the zero-field hopping Hamiltonian without the Hartree term, and $\mu_0$ is the matched bare chemical potential defined in Eq.~\eqref{app_mu_matching}. In this basis, the susceptibility is (see Appendix~\ref{app:Kab_derivation})
\begin{align}
\chi_{ij}(T) &= \sum_{a,b} K_{ab}(T) \phi_a(i)\phi_b(i) \phi_a^*(j)\phi_b^*(j),
\label{app_full_pair_kernel} \\
K_{ab}(T) &= \frac{1-f(\xi_a)-f(\xi_b)}{\xi_a+\xi_b} \nonumber\\
&= \frac{\tanh(\xi_a/2T)+\tanh(\xi_b/2T)} {2(\xi_a+\xi_b)} . \label{app_Kab}
\end{align}
The apparent pole at $\xi_a+\xi_b=0$ is removable at finite temperature:
\begin{equation}
K_{ab}(T) \rightarrow \frac{1}{4T} \operatorname{sech}^2 \left(\frac{\xi_a}{2T}\right).
\label{app_Kab_limit}
\end{equation}
In the numerical evaluation, Eq.~\eqref{app_Kab_limit} is used when $|\xi_a+\xi_b|\leq 10^{-12}t$; otherwise, the hyperbolic-tangent expression in Eq.~\eqref{app_Kab} is evaluated directly.

For the mode analysis, we restrict the Cooper susceptibility to the site set $S=\{i:0.03\le r_i/R_{\max}\le0.97\}$, where $R_{\max}=\max_i r_i\simeq69.805$ is the maximum radius of the Ti-site sample. The retained set contains $5660$ sites. The unrestricted susceptibility exhibits modes localized near the open boundary, where the lattice termination reduces the coordination number. To characterize the pairing structure of the interior quasiperiodic network, we therefore restrict the pairing-field subspace to $S$, while retaining the normal-state eigenstates of the full lattice.
We diagonalize this restricted susceptibility as
\begin{equation}
\sum_j \chi_{ij}(T)v_j^{(m)} = \lambda_m(T)v_i^{(m)}, \quad \lambda_1\geq \lambda_2\geq\cdots .
\label{app_Cij_eigenproblem}
\end{equation}
For this bare linearized problem, the $m$th channel reaches its instability criterion when $|U|\lambda_m(T)=1$. The eigenmodes and the temperature dependence of the leading eigenvalues are discussed in Appendix~\ref{app:origin_multifractal_pairing}.

\section{Derivation of the Cooper Response Factor}
\label{app:Kab_derivation}

This appendix derives the finite-temperature Cooper response factor $K_{ab}(T)$ appearing in Eq.~\eqref{app_full_pair_kernel}. The normal-state eigenbasis is defined by
\begin{equation}
c_{i\sigma}=\sum_a \phi_a(i)d_{a\sigma}, \qquad d_{a\sigma}=\sum_i \phi_a^*(i)c_{i\sigma}.
\label{app_cd_transform}
\end{equation}
The weak onsite pair source $\eta_i$ introduced in the preceding appendix is written in the normal-state eigenbasis as
\begin{align}
H_\eta &= -\sum_i
\left(\eta_i c_{i\uparrow}^{\dagger}c_{i\downarrow}^{\dagger}
+ \eta_i^*c_{i\downarrow}c_{i\uparrow}\right)
\nonumber\\
&= -\sum_{a,b} \left( \eta_{ab}d_{a\uparrow}^{\dagger}d_{b\downarrow}^{\dagger}
+ \eta_{ab}^*d_{b\downarrow}d_{a\uparrow} \right)
\label{app_H_delta_eigenbasis}
\end{align}
with
\begin{equation}
\eta_{ab} = \sum_i\eta_i\phi_a^*(i)\phi_b^*(i),
\qquad
F_{ab}=\langle d_{b\downarrow}d_{a\uparrow}\rangle\label{app_Delta_ab_Fab}
\end{equation}
The real-space anomalous amplitude is
\begin{equation}
F_i=\sum_{a,b}\phi_a(i)\phi_b(i)F_{ab}.
\label{app_Fi_from_Fab}
\end{equation}

Because the response is linear, one source matrix element $\eta_{ab}$ may be considered at a time. For a fixed orbital pair $(a,b)$, introduce
\begin{equation}
\Psi_{ab}=\begin{pmatrix} d_{a\uparrow}\\ d_{b\downarrow}^{\dagger} \end{pmatrix}, \qquad
\mathcal H_{ab} = \begin{pmatrix} \xi_a & -\eta_{ab}\\ -\eta_{ab}^* & -\xi_b \end{pmatrix}.\label{app_Nambu_block}
\end{equation}
The Matsubara Green's function is
\begin{equation}
\mathcal G_{ab}(i\omega_n)=\left[i\omega_n I-\mathcal H_{ab}\right]^{-1}, \qquad \omega_n=(2n+1)\pi T.
\label{app_Nambu_green}
\end{equation}
To linear order in $\eta_{ab}$, the anomalous Nambu component is
\begin{equation}
\left[\mathcal G_{ab}(i\omega_n)\right]_{12} = -\frac{\eta_{ab}} {(i\omega_n-\xi_a)(i\omega_n+\xi_b)} +O(\eta^3).\label{app_G12_linear}
\end{equation}
Here, the subscript $12$ denotes the off-diagonal anomalous component in Nambu space, while $a$ and $b$ label the normal-state orbitals in the two-particle Cooper channel. Thus
\begin{equation}
F_{ab}=T\sum_{\omega_n}e^{-i\omega_n0^-}\left[\mathcal G_{ab}(i\omega_n)\right]_{12}=-\eta_{ab}S_{ab},
\label{app_Fab_sum}
\end{equation}
where
\begin{equation}
S_{ab}=T\sum_{\omega_n}\frac{1}{(i\omega_n-\xi_a)(i\omega_n+\xi_b)}.
\label{app_Sab_def}
\end{equation}

Using partial fractions,
\begin{equation}
\frac{1}{(i\omega_n-\xi_a)(i\omega_n+\xi_b)}=\frac{1}{\xi_a+\xi_b}\left[\frac{1}{i\omega_n-\xi_a}-\frac{1}{i\omega_n+\xi_b}\right].
\label{app_partial_fraction}
\end{equation}
The symmetrically summed fermionic Matsubara identity
\begin{equation}
T\sum_{\omega_n}\frac{1}{i\omega_n-E}=f(E)-\frac{1}{2}
\label{app_fermion_sum}
\end{equation}
gives
\begin{align}
S_{ab}&=\frac{f(\xi_a)-f(-\xi_b)}{\xi_a+\xi_b}=-\frac{1-f(\xi_a)-f(\xi_b)}{\xi_a+\xi_b}.
\label{app_Sab_result}
\end{align}
The same summation prescription is used for both simple-pole terms in Eq.~\eqref{app_partial_fraction}, so their energy-independent constants cancel.
Combining Eqs.~\eqref{app_Fab_sum} and \eqref{app_Sab_result}, we obtain
\begin{equation}
F_{ab}=K_{ab}(T)\eta_{ab},
\label{app_Fab_Kab}
\end{equation}
where
\begin{align}
K_{ab}(T)&=\frac{1-f(\xi_a)-f(\xi_b)}{\xi_a+\xi_b} \nonumber\\
&=\frac{\tanh(\xi_a/2T)+\tanh(\xi_b/2T)}{2(\xi_a+\xi_b)} .
\label{app_Kab_derived}
\end{align}
When $\xi_a+\xi_b\to0$, Eq.~\eqref{app_Kab_derived} is understood through its smooth limiting value.

Substituting Eq.~\eqref{app_Fab_Kab} into Eq.~\eqref{app_Fi_from_Fab} yields
\begin{align}
F_i&=\sum_{a,b}\phi_a(i)\phi_b(i)K_{ab}(T)\sum_j \eta_j\phi_a^*(j)\phi_b^*(j) \nonumber\\
&=\sum_j \chi_{ij}(T)\eta_j,
\label{app_Fi_to_Cij}
\end{align}
with $\chi_{ij}(T)$ given by Eq.~\eqref{app_full_pair_kernel}. The pairing term in Eq.~\eqref{app_mean_field_hamiltonian} corresponds to $\eta_i=-\Delta_i$. Using $\Delta_i=UF_i$ with $U<0$, we obtain
\begin{equation}
\Delta_i = -U\sum_j\chi_{ij}(T)\Delta_j.
\label{app_linearized_gap_derived}
\end{equation}
where $U<0$ for the attractive onsite interaction.

\section{Normal-State Cooper Response and Hartree Feedback}
\label{app:origin_multifractal_pairing}

To identify a normal-state precursor of the multifractal pairing texture, we examine how critical near-Fermi eigenstates enter the bare Cooper susceptibility. On the Ti dual lattice, normal-state eigenfunctions have hierarchical amplitude distributions over the quasiperiodic geometry. The particle-particle form factor $\Phi_{ab}(i)=\phi_a(i)\phi_b(i)$ records where two energetically active eigenstates overlap locally, while $K_{ab}(T)$ weights each orbital pair by its finite-temperature Cooper response. The resulting $\chi_{ij}(T)$ is therefore a collective near-Fermi pair-overlap operator.

\begin{figure}
\includegraphics[width=0.95\columnwidth]{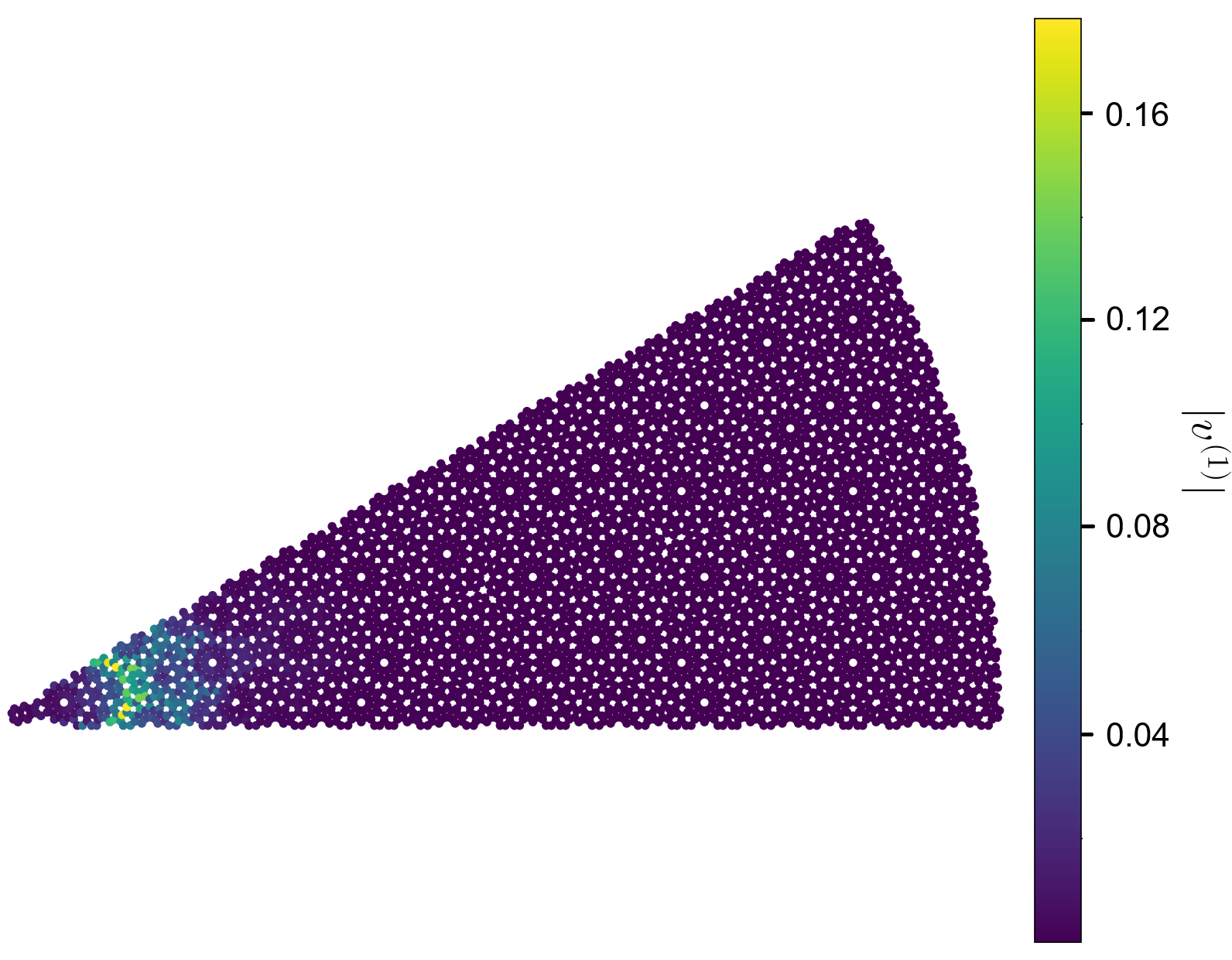}
\includegraphics[width=0.95\columnwidth]{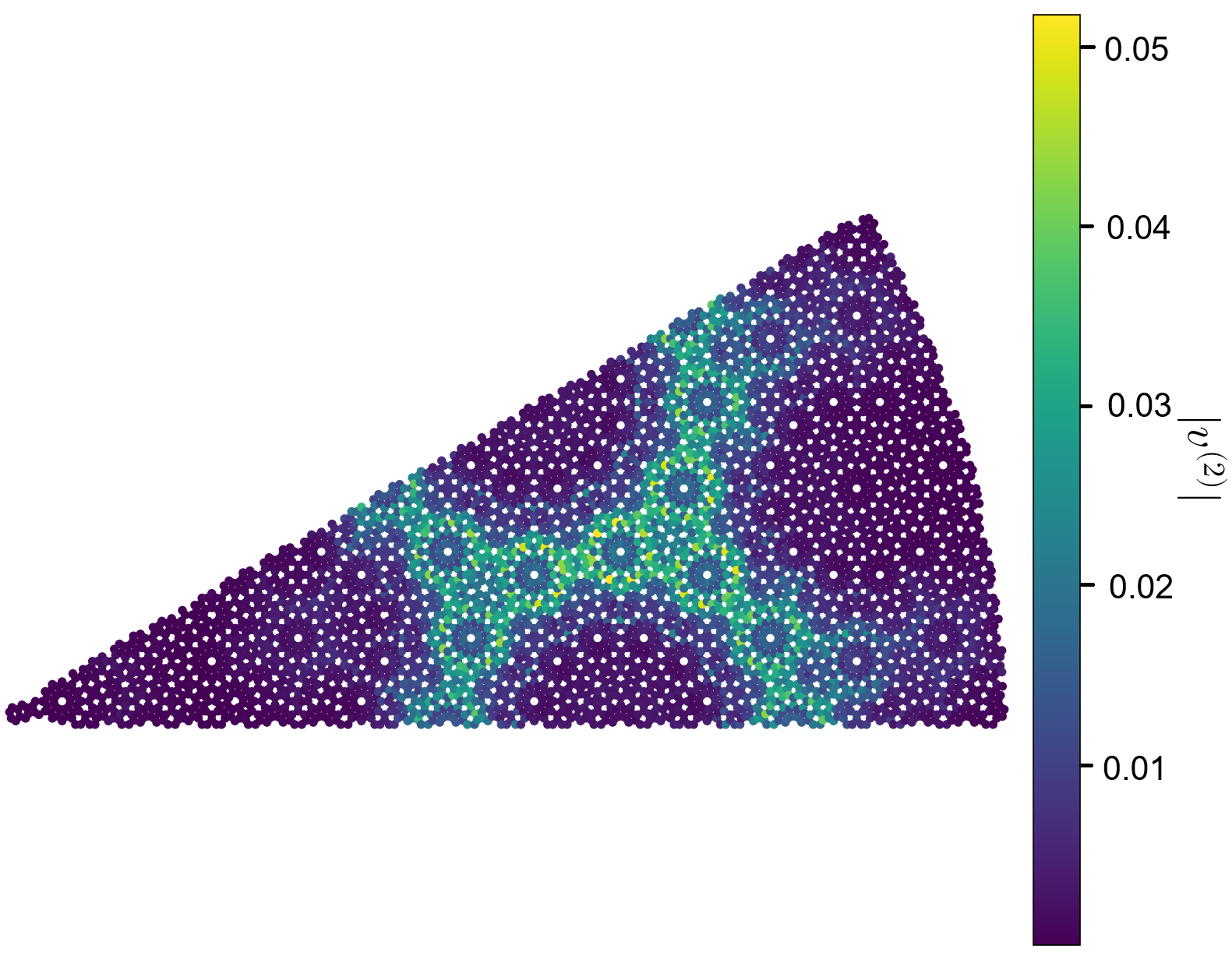}
\caption{
\label{fig:pair_kernel_mode_app}
Eigenmodes of the normal-state Cooper susceptibility kernel $\chi_{ij}$ at $T=0.015t$. The largest eigenmode $v^{(1)}$ is concentrated mainly near the innermost ring, whereas the leading extended mode $v^{(2)}$ exhibits the multi-generational ring-like structure observed in the self-consistent BdG order parameter.}
\end{figure}

\begin{figure}
\includegraphics[width=0.95\columnwidth]{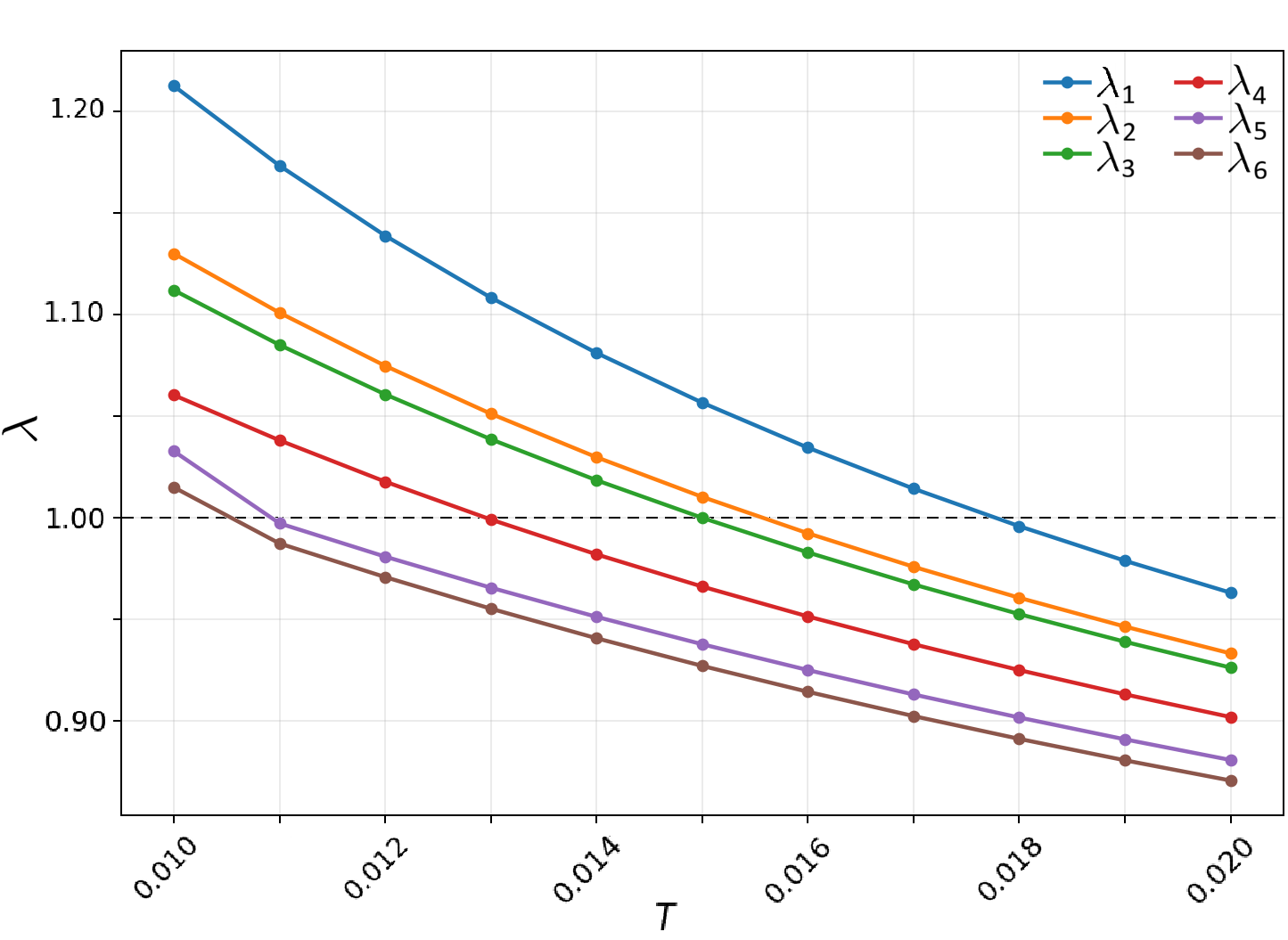}
\caption{
\label{fig:pair_kernel_eigenvalues_app}
Six largest eigenvalues $\lambda_m(T)$ of the normal-state Cooper susceptibility $\chi_{ij}$ as functions of temperature.}
\end{figure}

The eigenmodes of this operator are defined by Eq.~\eqref{app_Cij_eigenproblem}. In the finite system studied here, $v^{(1)}$ is localized mainly on the innermost ring. The mode $v^{(2)}$ has the largest eigenvalue among the modes extending over the larger hierarchical region, and its spatial support resembles the extended component of the self-consistent BdG order parameter. 
For eigenvectors normalized by
$\sum_{i\in\mathcal S}|v_i^{(m)}|^2=1$,
we define
\begin{equation}
w_S^{(m)}=\sum_{i\in S}|v_i^{(m)}|^2.
\end{equation}
At $T=0.015t$, the inner and extended weights are
$(w_{\rm inner}^{(1)},w_{\rm ext}^{(1)}) =(0.976,0.000183)$ and
$(w_{\rm inner}^{(2)},w_{\rm ext}^{(2)})=(0.00323,0.934)$.
These weights are not renormalized within $R$.
Figure~\ref{fig:pair_kernel_mode_app} shows the spatial structure of these susceptibility eigenmodes. Figure~\ref{fig:pair_kernel_eigenvalues_app} shows the temperature dependence of the leading eigenvalues; $|U|\lambda_1$ and $|U|\lambda_2$ reach unity at $T\simeq0.018\,t$ and $0.015\,t$, respectively. These crossings refer to the bare linearized problem and are not identified with the mean-field onset of the self-consistent Hartree--BdG solution.

The relation between the bare susceptibility $\chi_{ij}$ and the converged Hartree--BdG solution should be interpreted as a comparison between a linear-response diagnostic and a nonlinear self-consistent state. The bare susceptibility identifies spatial pairing channels with enhanced Cooper response. The self-consistent BdG calculation subsequently includes gap opening, Hartree shifts, spectral redistribution, and nonlinear mixing between response channels. The self-consistent pairing field can therefore reflect multiple response channels, with spatial weight in both the inner and extended regions. Thus $v^{(2)}$ and $|\Delta_i|$ need not agree site by site; the relevant comparison is whether they share the geometry-selected extended support. The observed correspondence therefore shows that an extended hierarchical component is already present in the bare normal-state response and remains visible after self-consistency.

The mode labels and the geometric-domain label play different roles. The eigenmodes $v^{(1)}$ and $v^{(2)}$ distinguish the compact and extended components of the Cooper spectrum, respectively, whereas $R$ is a mode-independent real-space domain. Because $R$ is fixed before $\chi_{ij}$ is computed (Appendix~\ref{app:second_radius_backbone}), the correspondence between $v^{(2)}$ and the extended component of $R$ is an outcome of the calculation, not an input.

The full Hartree--BdG calculation additionally includes the self-consistent Hartree field, which acts as a density-dependent local chemical-potential shift and modifies the quasiparticle spectrum during iteration. Because the bare susceptibility $\chi_{ij}$ is constructed without $V_i^{\rm H}$, the presence of the hierarchical extended response in $\chi_{ij}$ shows that this tendency already exists before Hartree feedback. However, the comparison does not by itself isolate the quantitative effect of the Hartree field on the final $|\Delta_i|$; doing so would require either a Hartree-free BdG comparison or a Hartree-dressed susceptibility constructed from the self-consistent $\Delta_i=0$ normal state.

\section{Geometry-defined second-radius domain}
\label{app:second_radius_backbone}

The centers used in the construction are not Ti sites. They are local void centers around which the neighboring Ti sites form approximately dodecagonal shells. Candidate centers are selected from their radial uniformity and 12-fold angular order without using the tight-binding spectrum, the BdG order parameter, or the Cooper kernel.

For the twelve nearest Ti sites around a candidate center, we evaluate the radial coefficient of variation $c_r=\operatorname{std}(\rho_j)/\overline{\rho}$ and the angular order
\begin{equation}
A_{12}=\left|\frac{1}{12} \sum_{j=1}^{12}e^{12i\theta_j}\right|.
\end{equation}
Candidates satisfying $c_r<10^{-3}$ and $A_{12}>0.999$ are retained.

\begin{figure}
\includegraphics[width=0.95\columnwidth]{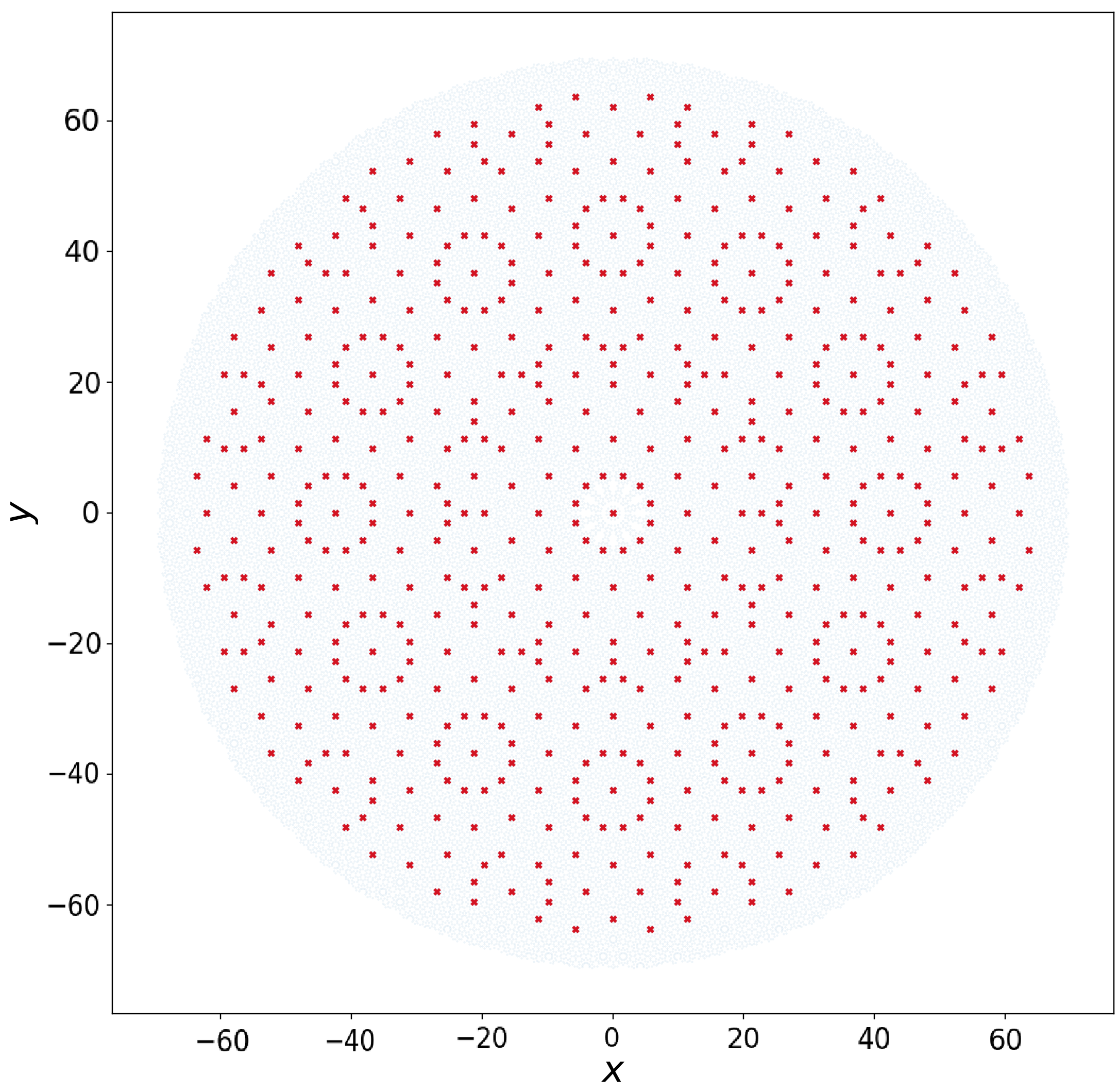}
\caption{\label{fig:symmetric_centers_app}
Local void centers with approximate 12-fold symmetry used to define the second-radius domain. The centers are geometric objects of the Ti-site network rather than Ti sites.}
\end{figure}

For two void centers $\mathbf R_\alpha$ and $\mathbf R_\beta$, equal circular domains first touch when their radius is one half of the center-to-center distance,
\begin{equation}
r_{\alpha\beta}=\frac{1}{2}|\mathbf R_\alpha-\mathbf R_\beta| .
\end{equation}
The distinct half-distances define a geometric hierarchy of contact radii. The first two relevant scales in the present finite sample are
\begin{equation}
r_1=\frac{2+\sqrt{3}}{\sqrt{6}} \simeq1.5236,\quad
r_2=\frac{9+5\sqrt{3}}{6} \simeq2.9434.
\end{equation}
Here $r_1$ is the smallest contact scale and $r_2$ is the next distinct scale. The value of $r_2$ is fixed by the center geometry rather than by fitting an electronic eigenmode. We apply the hard-core and boundary-contact criteria at this radius (retained $r_2$ disks do not overlap, and each touches at least one other retained disk) and retain the corresponding contact structure. The $R$ core-disk domain used below is the union of Ti sites inside the nominal $r_2$ disks of the retained centers, with the isolated central disk excluded. Its complement is denoted by $B$. Both $R$ and $B$ are therefore fixed before evaluating the electronic diagnostics.

\begin{figure}
\includegraphics[width=0.95\columnwidth]{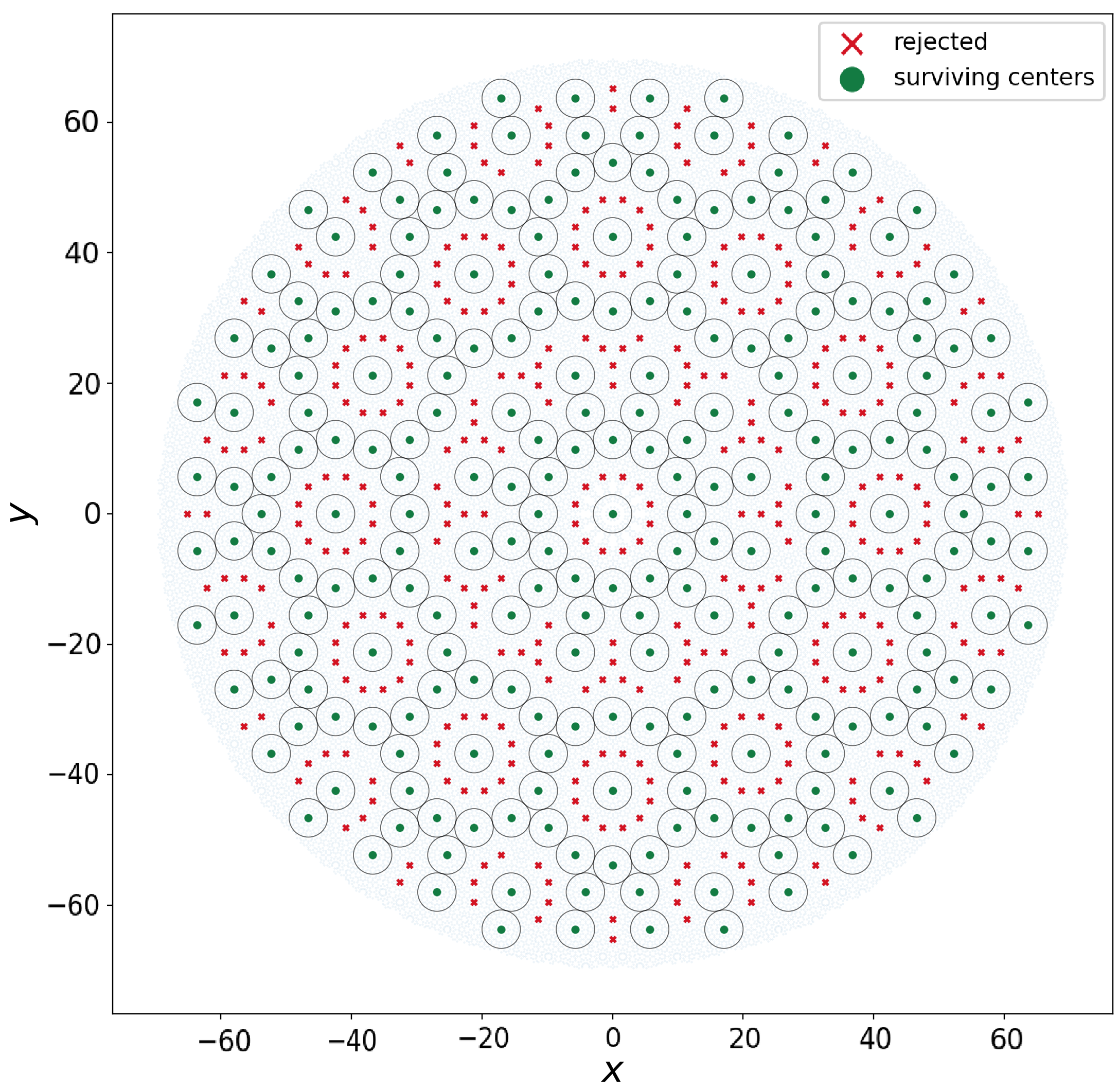}
\caption{\label{fig:r2_distribution_app}
Geometry-defined contact structure at the second radius $r_2$. Edges connect retained centers whose nominal $r_2$ disks meet at their boundaries.}
\end{figure}

\begin{figure}
\includegraphics[width=0.95\columnwidth]{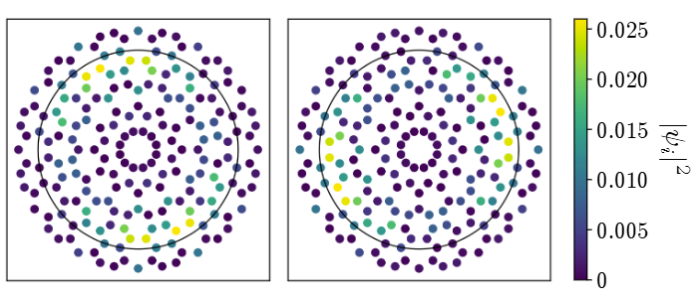}
\caption{\label{fig:local_r2_spectrum_app}
Representative local tight-binding eigenfunctions near $\mu_0=0.4t$ in a circular patch centered at a void center. The annular guide is centered at $r_2$ and has width $\delta r=0.6$.}
\end{figure}

As a local electronic illustration, we diagonalize a small tight-binding patch centered at a representative void center. A state near the bare normal-state chemical potential $\mu_0=0.4t$ has substantial probability near the $r_2$ boundary. The annular guide of width $\delta r=0.6$ in Fig.~\ref{fig:local_r2_spectrum_app} is used only to display this boundary and does not enter the definition of $R$.

\begin{figure}
\includegraphics[width=0.95\columnwidth]{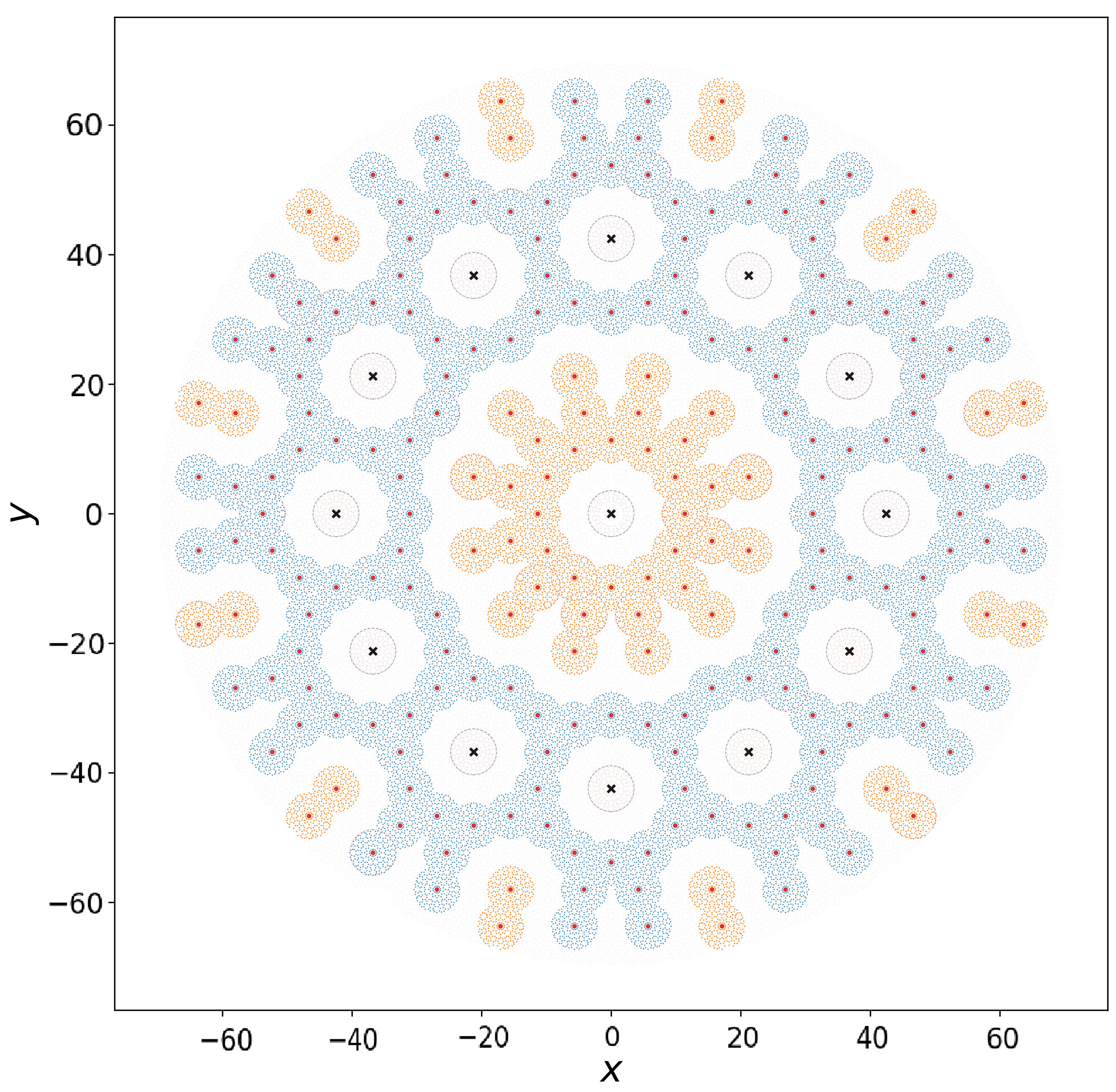}
\caption{\label{fig:R2_core_domain_app}
Geometry-defined domain $R$. The isolated central disk is excluded, and $B$ denotes the complementary site set. Blue and inner red distinguish the two connected components of $R$: the extended component and the smaller inner component.}
\end{figure}

\section{Multifractal Analysis}
\label{app:multifractal_spectrum}

We characterize the spatial distribution of the self-consistent order parameter by the normalized positive measure \begin{equation} p_i^{(\Delta)}=\frac{|\Delta_i|}{\sum_j|\Delta_j|}. \label{app_measure_delta} \end{equation} For the extended Cooper-susceptibility mode, the corresponding measure is \begin{equation} p_i^{(C)}=\frac{|v_i|}{\sum_j|v_j|}. \label{app_measure_C} \end{equation} The results reported below are obtained from the site-resolved $|\Delta_i|$ data. The same analysis pipeline must be applied separately to $p_i^{(C)}$ before making a quantitative comparison with the Cooper mode. Values satisfying $|\Delta_i|<10^{-6}t$ are treated as numerical zeros before normalization. No nonzero site or occupied cell is subsequently removed by a scale-dependent probability cutoff. The spatial measures are analyzed on the $30^\circ$ computational sector with $R_c=70$.

To avoid partial Cartesian boxes at the circular boundary, we use a boundary-conforming partition. At resolution $n$, the disk is divided into $n$ concentric annuli of equal radial width \begin{equation} \ell=\frac{R_c}{n},\qquad \epsilon=\frac{\ell}{R_c}=\frac{1}{n}. \label{app_dimensionless_box_size} \end{equation} The $k$th annulus is divided into \begin{equation} m_k=\operatorname{round}[2\pi(k+1/2)] \label{app_angular_cell_count} \end{equation} equal-angle cells. Their radial and tangential dimensions are both of order $\ell$, their areas are approximately $\ell^2$, and the outer edge coincides exactly with $r=R_c$. The mass in an occupied cell is \begin{equation} P_b(\epsilon) =\sum_{i\in b}p_i^{(\Delta)}. \label{app_box_measure} \end{equation} At each scale we use ten fractional angular shifts $\eta_\phi=(\phi-1)/10$, $\phi=1,\ldots,10$; in every annulus, $\eta_\phi$ is measured in units of that annulus's angular-cell width. The shifts leave the radial boundaries, angular cell counts, and total normalized mass unchanged, while individual cell assignments and masses are recomputed for every shift.

As a check of the box-shape dependence, we repeated the $q=2$ scaling analysis using Cartesian square boxes of side $\ell$, retaining only boxes whose four corners lie inside the circular sample and averaging over ten grid translations. Over the same scale window, this gives $D_2=1.944$ after normalizing the mass retained in the interior boxes, compared with $D_2=1.885$ for the polar partition. The small difference confirms that the estimate $D_2\simeq1.9$ is not controlled by the distorted innermost polar cells. We retain the polar partition for the full spectrum because it conserves the complete measure at every scale, whereas the fully contained square boxes discard a scale-dependent fraction of the boundary mass.

For each angular origin $\phi$, the partition sum is \begin{equation} Z_q^{(\phi)}(\epsilon) =\sum_{b:P_b^{(\phi)}>0}[P_b^{(\phi)}(\epsilon)]^q . \label{app_partition_sum} \end{equation} We average after taking the logarithm, \begin{equation} \overline{\ln Z_q(\epsilon)} =\frac{1}{N_\phi}\sum_{\phi=1}^{N_\phi} \ln Z_q^{(\phi)}(\epsilon), \label{app_log_partition_average} \end{equation} and extract the mass exponent from \begin{equation} \overline{\ln Z_q(\epsilon)} =\tau(q)\ln\epsilon+c_q. \label{app_partition_sum_scaling} \end{equation} The logarithmic average in Eq.~\eqref{app_log_partition_average}, rather than $\ln\overline{Z_q}$, is used consistently for all partition origins. For $q=1$, $Z_1^{(\phi)}=1$ for every origin, so $\overline{\ln Z_1}=0$ and $\tau(1)=0$.

For every occupied box and angular origin, the Chhabra--Jensen escort measure is \begin{equation} w_b^{(\phi)}(q,\epsilon)= \frac{[P_b^{(\phi)}(\epsilon)]^q} {\sum_{b'}[P_{b'}^{(\phi)}(\epsilon)]^q}. \label{app_escort_weight} \end{equation} The $q=0$ case is included by setting $w_b^{(\phi)}(0,\epsilon)=1/N_{\mathrm{occ}}^{(\phi)}(\epsilon)$. At each scale, the two Chhabra--Jensen logarithmic ordinates are evaluated separately for all ten angular origins and then averaged. The finite-sample spectrum is obtained from the regression slopes \begin{align} \frac{1}{N_\phi}\sum_\phi\sum_b w_b^{(\phi)}(q,\epsilon)\ln P_b^{(\phi)}(\epsilon) &=\alpha(q)\ln\epsilon+c_\alpha(q),\label{app_alpha_fit}\\ \frac{1}{N_\phi}\sum_\phi\sum_b w_b^{(\phi)}(q,\epsilon)\ln w_b^{(\phi)}(q,\epsilon) &=f(q)\ln\epsilon+c_f(q). \label{app_f_fit} \end{align} Here $f(q)\equiv f[\alpha(q)]$ parametrizes the plotted multifractal spectrum. This construction is equivalent to the Legendre transform $\alpha=d\tau/dq$, $f=q\alpha-\tau$ used in the main text, but avoids numerical differentiation of $\tau(q)$. We evaluate the escort measure over interval $-2.0\le q\le3.2$ in steps of $0.2$, using log-space exponentiation. Negative moments amplify the smallest box masses, whereas large positive moments concentrate on the largest box masses. This $q$ interval is for only $p_i^{(\Delta)}$, and both $p_i^{(\Delta)}$ and $p_i^{(C)}$ need interval $-1.2\le q \le 3.2$.

\begin{figure} \centering \includegraphics[width=0.98\columnwidth]{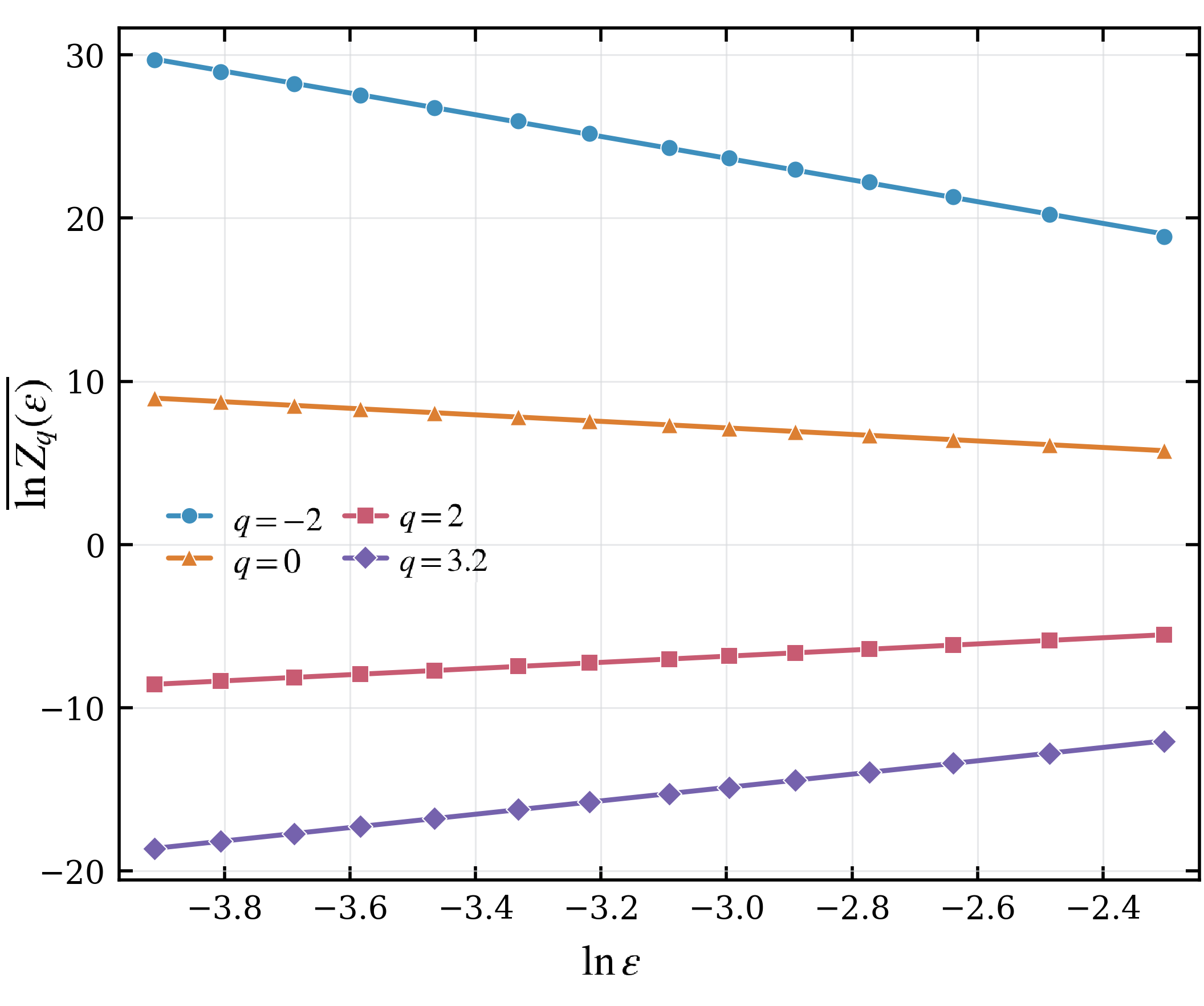} \caption{\label{fig:partition_sum_scaling_app} Partition-sum box-size scaling of the normalized order-parameter measure. Markers show the calculated $\overline{\ln Z_q(\epsilon)}$ values, and solid lines show linear fits over $0.02\le\epsilon\le0.10$ for representative moments within the accepted range.} \end{figure}

\begin{figure} \centering \includegraphics[width=0.95\columnwidth]{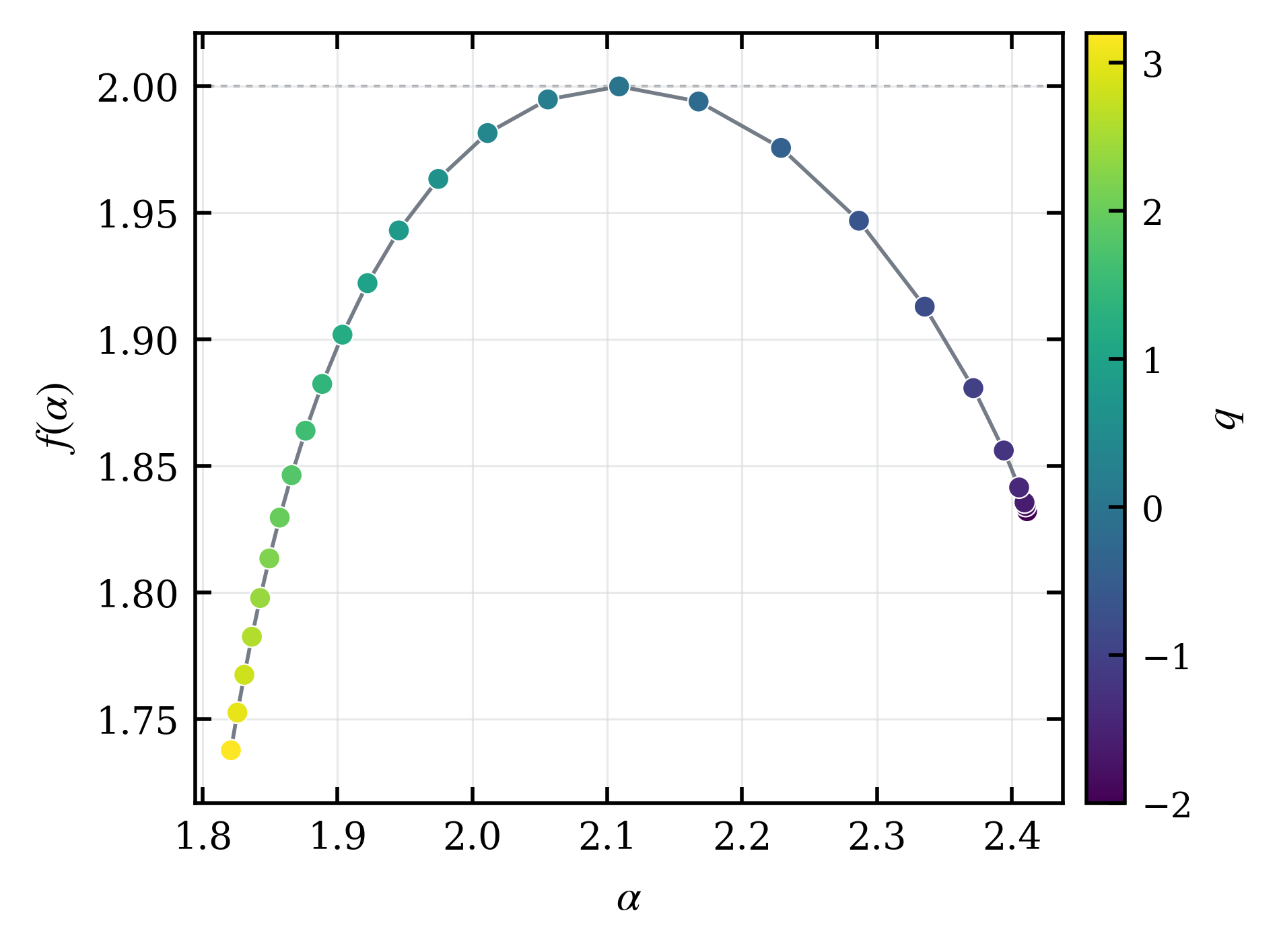} \caption{\label{fig:multifractal_spectrum_app} Chhabra--Jensen spectrum of the normalized order-parameter amplitude $|\Delta_i|$. The $|\Delta_i|$ spectrum is evaluated on the full circular sample, while the eigenmode spectrum uses the available annular support $0.03<r/R_c<0.97$. Only the stable moment ranges $-2.0\le q\le3.2$, respectively, are shown; markers denote sampled moments and lines connect adjacent values.} \end{figure}

The analysis uses the common intermediate-scale window $0.02\le\epsilon\le0.10$, excluding finer scales affected by discrete-site saturation and coarser scales showing curvature. Varying $\epsilon_{\min}$ and $\epsilon_{\max}$ independently by $\pm20\%$ from $0.02$ and $0.10$, respectively, gives $D_2=1.871\text{--}1.890$, within $0.8\%$ of the original value $1.885$. Representative linear scaling of $\overline{\ln Z_q}$ with $\ln\epsilon$ is shown in Fig.~\ref{fig:partition_sum_scaling_app}. Only stable moment intervals are reported, without applying a scale-dependent probability cutoff, smoothing, or parabolic extrapolation. Since the box size is varied within a single finite sample, the resulting spectra represent resolved finite-sample scaling rather than finite-size scaling toward the thermodynamic limit. So the reported exponents refer to the specified partition and intermediate-scale fitting window.

\section{Supercurrent Under Applied Flux}
\label{app:supercurrent}

To calculate the equilibrium current response, we couple the Ti-site network to an external vector potential through the Peierls substitution
\begin{equation}
t_{ij}\rightarrow t_{ij}e^{i\phi_{ij}}, \qquad
\phi_{ij}=-\frac{e}{\hbar}\int_{\mathbf r_j}^{\mathbf r_i}\mathbf A\cdot d\mathbf l .
\label{app_peierls_phase}
\end{equation}
Here, $e>0$ is the elementary charge, so the electron charge is $-e$. We set $e=\hbar=1$ in the numerical calculations after fixing the Peierls-phase convention.

For the current pattern in Fig.~\ref{fig:main}(d), we use the circulating vector potential
\begin{equation}
\mathbf A_{\Phi}(x,y) = 3.3 \left( -\frac{y}{x^2+y^2}, \frac{x}{x^2+y^2} \right),
\label{app_flux_vector_potential}
\end{equation}
centered at the sample origin. This calculation is performed on a smaller circular sample of $N_s=7656$ Ti sites, constructed as in Appendix~\ref{app:construction} with a smaller cutoff radius and centered on the same 12-fold rotation axis, without the $30^\circ$ sector reduction, since the sector identification would restrict the phase winding around the flux line to multiples of 12. The external vector potential is held fixed while both the complex pairing field $\Delta_i$ and the Hartree potential $V_i^{\rm H}$ are determined self-consistently. The line integral is evaluated along each straight Ti--Ti bond using the endpoint angular difference wrapped to $(-\pi,\pi]$, with $\phi_{ji}=-\phi_{ij}$.

The bond current from site $j$ to site $i$ is obtained from the derivative of the Hamiltonian with respect to the Peierls phase,
\begin{equation}
I_{j\to i} = \frac{2e}{\hbar} \operatorname{Im} \left[ t_{ij}e^{i\phi_{ij}} \sum_{\sigma} \left\langle c_{i\sigma}^{\dagger}c_{j\sigma} \right\rangle \right].
\label{app_bond_current}
\end{equation}
For a spin-balanced BdG solution, the spin-summed density matrix is
\begin{equation}
\begin{aligned}
\sum_{\sigma} \left\langle c_{i\sigma}^{\dagger}c_{j\sigma} \right\rangle
= 2\sum_{E_n>0}
[ u_n^*(i)u_n(j)f(E_n) \\
+ v_n(i)v_n^*(j) \{1-f(E_n)\} ].
\label{app_density_matrix_bdg}
\end{aligned}
\end{equation}
The vector current plotted in Fig.~\ref{fig:main}(d) is obtained by assigning the scalar current to the corresponding bond direction,
\begin{equation}
\mathbf J_{j\to i} = I_{j\to i} \frac{\mathbf r_i-\mathbf r_j} {|\mathbf r_i-\mathbf r_j|}.
\label{app_vector_current}
\end{equation}
This is the total equilibrium BdG bond current in the presence of the applied vector potential. At the same flux and parameters, the equilibrium current vanishes when $\Delta_i=0$, so no normal-state current is subtracted.

To make the phase response explicit, we write the local superconducting order parameter as
\begin{equation}
\Delta_i = |\Delta_i|e^{i\theta_i},
\label{app_order_parameter_phase}
\end{equation}
where $\theta_i$ is the phase on Ti site $i$. With the convention in Eq.~\eqref{app_peierls_phase}, the gauge-invariant phase twist on a directed bond $j\to i$ is
\begin{equation}
\Theta_{ij} = \theta_i-\theta_j-2\phi_{ij}.
\label{app_gauge_invariant_phase_twist}
\end{equation}
The factor of two reflects the charge $-2e$ of a Cooper pair. The bond response is therefore controlled by the gauge-invariant combination of the order-parameter phase and the Peierls phase, rather than by either contribution separately.

For a closed contour $\mathcal C$ along which $|\Delta_i|$ remains nonzero, single-valuedness of the pair field requires
\begin{equation}
\sum_{(j\to i)\in\mathcal C} (\theta_i-\theta_j) = 2\pi m, \qquad m\in\mathbb Z .
\label{app_phase_winding_discrete}
\end{equation}
The corresponding superconducting flux quantum is
\begin{equation}
\Phi_0 = \frac{h}{2e}.
\label{app_flux_quantum}
\end{equation}
For a contour enclosing the flux core, the applied Peierls phases generally cannot be canceled on every bond by a single-valued choice of $\theta_i$. The self-consistent solution then supports circulating equilibrium-current paths.

The current distribution in Fig.~\ref{fig:main}(d) shows that these paths predominantly follow the ring structures on which $|\Delta_i|$ is concentrated. It therefore demonstrates how the finite-flux equilibrium response is spatially organized by the multifractal pairing pattern within the self-consistent BdG solution. Because this response is evaluated at a fixed finite flux and contains both condensate and quasiparticle contributions, it does not by itself determine the phase stiffness, which we calculate independently from the weak-probe response of the grand potential in Appendix~\ref{app:bkt_estimate}.

\section{Superfluid Phase Stiffness}
\label{app:bkt_estimate}

Using the BdG procedure of Appendix~\ref{app:supercurrent}, we compare two low-temperature responses of the same finite sample. The complex pairing field and Hartree densities are determined self-consistently, subject to the phase constraints specified below.

\begin{table}[!t]
\caption{\label{tab:stiffness_app}
Mean-field responses for the orbital probe and the zero-field twist ($w=1.0$, $d_i\geq1.5$).}
\begin{ruledtabular}
\begin{tabular}{ccc}
$T/t$ & $Y_{\rm orb}/(10^{-3}t)$ & $Y_{\rm tw}/(10^{-3}t)$\\
\hline
0.002 & 5.652 & 4.617\\
0.006 & 5.044 & 4.190\\
0.007 & 4.581 & 3.850\\
0.008 & 3.991 & 3.409\\
\end{tabular}
\end{ruledtabular}
\end{table}

The orbital probe is
\begin{align}
\mathbf A_a&=a(-y,x),\qquad
\nabla\times\mathbf A_a=2a\hat{\mathbf z},
\label{app_stiffness_vector_potential}\\
\phi_{ij}^{(a)}&=a(x_i y_j-x_j y_i).
\label{app_stiffness_bond_phase}
\end{align}
On the zero-winding branch continued from $a=0$, we write
\begin{equation}
\delta\Omega(a)\equiv\Omega(a)-\Omega(0)
=\tfrac12c_2a^2+O(a^4),
\label{app_grand_potential_curvature}
\end{equation}
and use Richardson extrapolation with $\delta a=10^{-3}$:
\begin{equation}
c_2\simeq
\frac{16\delta\Omega(\delta a)-\delta\Omega(2\delta a)}
{6(\delta a)^2}.
\label{app_stiffness_c2}
\end{equation}
The orbital response is normalized as
\begin{equation}
Y_{\rm orb}=\frac{c_2}{4\mathcal A_{\rm hull}\langle r^2\rangle},
\qquad
\langle r^2\rangle=N^{-1}\sum_i r_i^2,
\label{app_stiffness_normalization}
\end{equation}
where $\mathcal A_{\rm hull}=1600.96$ and $\langle r^2\rangle=257.44$. This uses the uniform-continuum normalization without imposing a uniform phase on the BdG solution.

At $T=0.002t$, reducing $\delta a$ to $5\times10^{-4}$ and $2.5\times10^{-4}$ changes $c_2$ by less than $0.004\%$. At $a=2\times10^{-3}$, a winding state has lower energy; the quoted curvature refers to the continued zero-winding branch. The constrained normal reference ($\Delta_i=0$), evaluated at its zero-field stationary Hartree density, gives $c_2^{\rm N}/c_2=-0.0172$. This contribution is retained in $Y_{\rm orb}$.

To probe the phase response without an applied magnetic field, we impose a zero-field twist, $\mathbf A=0$ and only the pairing phase is pinned to $\theta_i=qx_i$ on $d_i<w$, where $d_i$ is the distance to the convex-hull boundary. All amplitudes, Hartree densities, and unpinned phases remain self-consistent. We use $w=1.0$ and $qL_x=0.2$, with $L_x=x_{\max}-x_{\min}$. In the window $W=\{i:d_i\geq1.5\}$, with convex-hull area $\mathcal A_W$, we define the linear-response estimate
\begin{equation}
E_W=\tfrac12\sum_{b\subset W}I_b\chi_b,
\qquad
Y_{\rm tw}=\frac{2E_W}{\mathcal A_W(q')^2},
\label{app_twist_response}
\end{equation}
where each bond $b=(i,j)$ is counted once,
$I_b=\partial\Omega/\partial\phi_{ij}$ is the spin-summed current, and
$\chi_b=\operatorname{Arg}(\Delta_j\Delta_i^*)/2$.
The realized gradient $q'$ is fitted from
$\theta_i=q'x_i+c_y y_i+c_0$ on $1.5\leq d_i<2.0$.

At $T=0.002t$, the $w=0.5$ check gives
$\delta\Omega(2q)/\delta\Omega(q)=3.9956$; energy and current estimates of the curvature agree within $1.20\%$. Changing $w$ between $0.5$ and $1.0$ changes $Y_{\rm tw}$ by at most $1\%$. Varying the interior window yields $0.72$--$1.0\,Y_{\rm orb}$ without a plateau, so the reported twist response refers specifically to the stated window.

These probe-dependent responses remain finite at the sampled temperatures (Table~\ref{tab:stiffness_app}). Thermal vortex renormalization is not included, and no BKT transition temperature is inferred.

\end{document}